\documentclass[numberedappendix,twocolumn]{emulateapj}
\usepackage{graphicx}
\usepackage{natbib}
\usepackage{amsmath}
\usepackage{color}
\shorttitle{Magneto-optical effects in the wings of the Ca~{\sc i} 
4227~\AA\ line}
\shortauthors{Alsina Ballester, Belluzzi \& Trujillo Bueno}
\begin{document}
\title{Magneto-optical effects in the scattering polarization wings\\ 
of the Ca~{\sc i} $4227$~\AA\ resonance line.}
\author{E. Alsina Ballester\altaffilmark{1,2}, L. Belluzzi\altaffilmark{3,4} and J. Trujillo Bueno\altaffilmark{1,2,5}} \altaffiltext{1}{Instituto de Astrof\'{\i}sica de Canarias, E-38205 La Laguna, Tenerife, Spain}\altaffiltext{2}{Departamento de Astrof\'\i sica, Facultad de F\'\i sica, Universidad de La Laguna, E-38206 La Laguna, Tenerife, Spain}\altaffiltext{3}{Istituto Ricerche Solari Locarno, CH-6605 Locarno Monti, Switzerland}\altaffiltext{4}{Kiepenheuer-Institut f\"ur Sonnenphysik D-79104 Freiburg, Germany}\altaffiltext{5}{Consejo Superior de Investigaciones Cient\'{\i}ficas, Spain}
\email{ealsina@iac.es}
\begin{abstract}
The linear polarization pattern produced by scattering processes in the Ca~{\sc i} $4227$~\AA\ resonance line is a valuable observable for probing the solar atmosphere. Via the Hanle effect, the very significant $Q/I$ and $U/I$ line-center signals are sensitive to the presence of magnetic fields in the lower chromosphere with strengths between $5$ and $125$~G, approximately. On the other hand, partial frequency redistribution (PRD) produces sizable signals in the wings of the $Q/I$ profile, which have always been thought to be insensitive to the presence of magnetic fields. Interestingly, novel observations of this line revealed a surprising behavior: fully unexpected signals in the wings of the $U/I$ profile and spatial variability in the wings of both $Q/I$ and $U/I$. Here we show that the magneto-optical (MO) terms of the Stokes-vector transfer equation produce sizable signals in the wings of $U/I$ and a clear sensitivity of the $Q/I$ and $U/I$ wings to the presence of photospheric magnetic fields with strengths similar to those that produce the Hanle effect in the line core. This radiative transfer investigation on the joint action of scattering processes and the Hanle and Zeeman effects in the Ca~{\sc i} $4227$~\AA\ line should facilitate the development of more reliable techniques for exploring the magnetism of stellar atmospheres. To this end, we can now exploit the circular polarization produced by the Zeeman effect, the magnetic sensitivity caused by the above-mentioned MO effects in the $Q/I$ and $U/I$ wings, and the Hanle effect in the line core.
\end{abstract}
\newpage
\section{Introduction}
\label{Sect:Introd}
The Ca~{\sc i} line at $4227$~\AA\ presents one of the largest scattering 
polarization signals of the so-called Second Solar Spectrum 
\citep[e.g.,][]{Gandorfer02}, and indeed, it was among the first 
ones to be observed \citep[e.g.,][]{Bruckner63,Stenflo74,Stenflo82,Wiehr75}.
In the intensity spectrum, this line has a broad absorption profile 
with extended wings, while in the linearly-polarized spectrum due to scattering processes
(the Second Solar Spectrum) it shows a peculiar 
triplet peak structure $Q/I$ profile, with a sharp peak in the line core, 
and polarizing lobes in the wings. 
In observations of quiet regions close to the limb, the central peak typically has 
an amplitude of approximately 2\%, while the wing lobes reach
amplitudes of about $3$\%.
The core of this line is sensitive to the thermodynamic conditions in the lower 
chromosphere (its formation height is around $1000$~km above the 
$\tau_{5000} = 1$ surface). 
At such altitudes, the density of perturbers is relatively low and scattering
is basically coherent in the atomic rest frame. Coherent scattering with
partial frequency redistribution (PRD) due to the Doppler effect in the
observer's frame are key physical ingredients at the origin of the
aforementioned extended $Q/I$ wing signals \citep[e.g.,][]{Auer+80}. 
The linear polarization in the line core is modified through the Hanle effect
\citep[see][]{Faurobert92,Frisch96}. 
For this line the Hanle critical field\footnote{The Hanle critical field for
the onset of the Hanle effect is given by $B_c = 1.137 \cdot 10^{-7} 
\Gamma_R/g_u$, where $\Gamma_R$ is the radiative line broadening parameter and 
$g_u$ is the upper level's Land\'e factor.} is close to $25$~G, so we can
expect a noticeable impact on the line core amplitudes of $Q/I$ and $U/I$ for
field strengths as weak as $5$~G.

The Ca~{\sc i} $4227$~\AA\ line is produced by a resonance transition between
an upper level with total angular momentum $J_u = 1$, and a lower level with
$J_\ell = 0$. 
Moreover, its upper level is not radiatively coupled to any level with lower 
energy other than the lower level of the transition which, being the ground 
level, has a very long lifetime. Therefore, this spectral line can be suitably 
modeled using a two-level model atom with an infinitely sharp and unpolarized lower level.
 
\citet{Anusha+10} studied the non-magnetic behavior of the Ca~{\sc i}
$4227$~\AA\ line by considering various generalizations of the last scattering 
approximation. 
Later, \citet{Anusha+11} performed PRD radiative transfer (RT) 
calculations for the polarization profiles accounting for the Hanle effect 
together with the longitudinal Zeeman effect, with the aim of determining 
the magnetic field vector from disk-center observations. 
\citet{CarlinBianda17} studied the sensitivity of this line to the vertical gradients of the vertical component of the macroscopic velocity in a three-dimensional (3D) solar model atmosphere, neglecting 
the symmetry breaking effects caused by the horizontal transfer of radiation.

Spectropolarimetric observations of the Ca~{\sc i} $4227$~\AA\ line were 
conducted at IRSOL using ZIMPOL-II \citep[see][]{Bianda+03}. 
Such observations showed a surprising result, namely that the $U/I$ profiles can also 
be very significant in the line wings and that there are spatial variations of the wing signals, 
both in $Q/I$ and $U/I$. 
\citet{Sampoorna+09} also {presented} observations of this line with ZIMPOL-II, 
{pointing out a similar behavior}. 
Their study, based on the last scattering approximation, led them to discard the 
possibility that the Hanle effect could operate outside the line core region due to the impact 
of elastic collisions, and they suggested that the mysterious behavior of the $Q/I$ and $U/I$ wings 
could perhaps have a non-magnetic 
origin in terms of local inhomogeneities. 
In this paper, we present an alternative explanation in terms of RT effects 
that modify the radiation as it propagates through the magnetized solar 
atmosphere and introduce magnetic sensitivity in the wings of $Q/I$ and $U/I$. 

The fact that this line has a clear sensitivity to PRD effects, together with 
the fact that it can be modeled using the two-level atom model with an unpolarized 
and infinitely sharp lower level, makes it an ideal candidate to be studied 
through the forward modeling approach and numerical code presented in
\citet{Alsina+17}, hereafter ABT17. In such paper, as well as in \citet{Alsina+16}, 
we reported our theoretical discovery, achieved within the framework of the PhD thesis of the first author of this paper, that, in strong resonance lines for which the effects of PRD are important, the magneto-optical 
(MO) $\rho_V \, Q $ and $\rho_V \, U$ terms of the transfer equations for Stokes $U$ and $Q$, 
respectively, can produce significant $U/I$ wing signals and a magnetic sensitivity in the wings of 
both $Q/I$ and $U/I$. In ABT17 we illustrated the significance of this new polarization mechanism 
by solving the radiative transfer problem for the Sr~{\sc ii} $4078$~\AA\ line, while in 
\citet{Alsina+16} we demonstrated its impact on the near wings of the Mg~{\sc ii} $k$ line at 
$2795.5$~\AA\ (see also \citealt{delPino+16} {and} \citealt{MansoSainz+17} for a two-term model atom investigation of the observational signatures of such MO effects across the $h$ \& $k$ lines of Mg~{\sc ii}).

Motivated by the surprising spectropolarimetric observations of the Ca~{\sc i} $4227$~\AA\ line mentioned above, 
and by the potential interest of this resonance line for probing the solar atmosphere, we have carried out 
a detailed radiative transfer investigation about the impact of scattering processes and the joint 
action of the Hanle and Zeeman effects on the Stokes profiles of this line. Since the problem is 
very complex, given that we account for the effects of PRD in the presence of arbitrary magnetic 
fields, here we use one-dimensional (1D) semiempirical models of the solar atmosphere. This is 
sufficient to show clearly that the presence of weak magnetic fields in the region of formation 
of the line wings (i.e., the solar photosphere) can produce very significant signals in the wings 
of the $U/I$ profile, and a conspicuous magnetic sensitivity in both the $Q/I$ and $U/I$ wings.  

\section{Formulation of the problem}

We model the Ca~{\sc i} $4227$~{\AA} line using the RT code described in 
ABT17, which calculates the Stokes profiles of the emergent radiation 
assuming a two-level atom with an unpolarized and infinitely sharp lower 
level, and taking PRD effects into account, according to the theoretical 
approach of \citet{Bommier97a,Bommier97b}. 
The code accounts for the splitting of the magnetic sublevels produced by 
a magnetic field of arbitrary strength and orientation (Zeeman effect), as 
well as for its impact on the atomic level polarization (Hanle effect). 
Stimulated emission is neglected, which is a reasonable approximation when 
considering radiation fields weak enough that the average number of photons 
per mode is much smaller than unity, as is the case for the radiation fields 
found in the solar atmosphere. 

The intensity and polarization of a monochromatic beam, with frequency $\nu$ 
and propagating in direction $\vec{\Omega}$, are governed by the RT equations
\begin{equation}
\frac{\mathrm{d}}{\mathrm{d}s} I_i = \varepsilon_i - K_{i j} I_j \, ,
\label{RTE}
\end{equation}
where $s$ is the geometrical distance along $\vec{\Omega}$. 
In this equation the Stokes parameters $I_i$ with $i = [0,1,2,3] = [I,Q,U,V]$
are coupled to one another through the propagation matrix
\begin{equation}
 \hat{K} =  \left(\begin{array}{c c c c}
 \eta_I & \eta_Q & \eta_U & \eta_V \\
 \eta_Q & \eta_I  & \rho_V & -\rho_U \\
 \eta_U & -\rho_V & \eta_I & \rho_Q \\
 \eta_V & \rho_U & -\rho_Q & \eta_I \end{array} \right) \, .
  \label{Prop}
  \end{equation}
The quantity $\eta_I$ is the absorption coefficient, which quantifies the 
extinction of a radiation beam as it propagates through the atmospheric
material. The dichroism terms $\eta_Q$, $\eta_U$ and $\eta_V$ quantify the
differential absorption of radiation in a given polarization state, while the 
anomalous dispersion terms $\rho_Q$, $\rho_U$ and $\rho_V$ measure the 
coupling between different polarization states. 
From the explicit expression of these terms \citep[][hereafter LL04]{LandiLandolfi04}, it can be seen 
that if no lower level polarization is present (as it is the case of the line 
under investigation, which has $J_{\ell}=0$), the propagation matrix is 
diagonal unless a magnetic field is present and the Zeeman splitting is 
taken into account in the line profiles.
The quantities $\varepsilon_{i}$ are the emission coefficients for each Stokes 
parameter. 
In general, all the RT coefficients depend on the spatial point, frequency, 
and direction, and they can be decomposed into the sum of a line and a
continuum component. 
 
In the visible part of the solar spectrum, continuum processes do not bring 
any contribution to dichroism and anomalous dispersion (e.g., LL04), so that
the continuum part of the propagation matrix is diagonal.
The continuum emission coefficient has a thermal component, and a scattering 
 component, which we treat as purely coherent in the observer's reference
frame (see ABT17 for more details). 

For a two-level atom with an unpolarized lower level, the line components of 
the propagation matrix elements, considering a reference frame in which the 
quantization axis for total angular momentum is parallel to the magnetic field, 
are given by 
\begin{subequations}
\begin{align}
& \eta^\ell_i(\nu,\vec{\Omega}) = k_L \sum_K \Phi^{0, K}_0(J_\ell, J_u; \nu) {\mathcal T}^K_0(i,\vec{\Omega}) \, , \label{Eta2lev} \\
& \rho^\ell_i(\nu,\vec{\Omega}) = k_L \sum_K \Psi^{0, K}_0(J_\ell, J_u; \nu) {\mathcal T}^K_0(i,\vec{\Omega}) \, , \label{Rho2lev}
\end{align}
\label{Prop2lev}
\end{subequations}
where the apex $\ell$ indicates that the corresponding quantity represents 
the line contribution only. 
The polarization tensor ${\mathcal T}^K_Q(i,\vec{\Omega})$ is defined in 
Chapter 5 of LL04.
The generalized profile $\Phi^{K, K'}_Q(J_\ell,J_u;\nu)$ and the generalized 
dispersion profile $\Psi^{K, K'}_Q(J_\ell,J_u;\nu)$ are defined and discussed 
in detail in Appendix 13 of LL04, and can also be found in ABT17. 
 
The line contribution to the emission coefficient can be divided into a 
radiative and a thermal part
 \begin{equation}
 \varepsilon^\ell_i (\nu, \vec{\Omega}) = \varepsilon^{\ell,\mbox{\scriptsize th}}_i(\nu,\vec{\Omega}) + \varepsilon^{\ell, \mbox{\scriptsize rad}}_i(\nu,\vec{\Omega}) \, .
 \label{Em_line}
 \end{equation}
The expression of the thermal part can  
be found in ABT17. 
Working within the framework of the redistribution matrix formalism 
\citep[e.g.,][]{Hummer62}, the radiative part is given by
 \begin{align}
 \varepsilon^{\ell, \mbox{\scriptsize rad}}_i & (\nu, \vec{\Omega}) = k_L \int \! \mathrm{d}\nu' \oint\ \! \frac{\mathrm{d}\vec{\Omega}'}{4 \pi} \,
 \sum_{j=0}^3 \bigl[{\mathcal R}(\nu',\vec{\Omega}',\nu,\vec{\Omega};\vec{B}) \bigr]_{ij} \notag \\
 & \times  I _{j}(\nu',\vec{\Omega'}) \, .
\label{Eml_rd} 
 \end{align}
The convention according to which primed quantities refer to the incident
radiation, while unprimed ones refer to the scattered radiation has been used.
The redistribution matrix $\bigl[{\mathcal R}(\nu',\vec{\Omega}',\nu,
\vec{\Omega};\vec{B}) \bigr]_{ij}$ relates the frequency, direction, and
polarization properties of the incoming radiation to those of the scattered
radiation. 
For the case of a two-level atom with an infinitely sharp lower level, it can
be expressed as a linear combination of two terms \citep{Hummer62} 
 \begin{align}
 \bigl[{\mathcal R} & (\nu',\vec{\Omega}',\nu,\vec{\Omega};\vec{B}) \bigr]_{ij} =
 \bigl[{\mathcal R}_{\mbox{\sc ii}}(\nu',\vec{\Omega}',\nu,\vec{\Omega};\vec{B}) \bigr]_{ij}  \notag \\
  &+ \bigl[{\mathcal R}_{\mbox{\sc iii}}(\nu',\vec{\Omega}',\nu,\vec{\Omega};\vec{B}) \bigr]_{ij} \, .
 \label{RedisSep}
 \end{align}
The first term on the rhs represents the contribution due to scattering 
processes which are coherent in the atomic rest frame, while the second one 
represents the contribution due to scattering processes in which the
frequencies of the incoming and scattered radiation are completely 
uncorrelated (CRD). 
The explicit expressions for these contributions, originally derived 
by \cite{Bommier97b}, are written in ABT17, taking into account Doppler 
redistribution in the observer's reference frame, and taking the quantization 
axis for total angular momentum along an arbitrary direction. 
Therein, the iterative method applied in order to obtain a converged solution 
of the full non-LTE RT problem is also detailed. 
 
The calculations presented in this paper are carried out considering the 
semiempirical models of \cite{Fontenla+93} and \cite{Avrett95}. 
The non-LTE RT problem is first solved neglecting polarization, using 
the Multi-Level Accelerated Lambda Iteration (MALI) code developed by 
\cite{Uitenbroek01}, which we will hereafter refer to as RH, in the absence of 
magnetic fields. 
This RT code is executed considering a Ca~{\sc i} atomic model which consists 
of 20 levels, including the ground level of Ca~{\sc ii}, and accounts for 17 
line transitions and 19 continuum transitions.  
All spectral lines are computed assuming CRD, except for the $4227$~\AA\ line 
itself, which receives a PRD treatment. 
This calculation provides the population of the lower level (which will then be 
fixed when solving the non-LTE polarized RT problem), as well as the initial 
estimate of the radiation field present at each height point in the 
atmosphere. 
The RH code is also used to calculate the values of the collisional rates, 
and of the continuum thermal emissivity, which are then required as
inputs for the code described in ABT17.  

\begin{figure}[!ht]
\centering
\includegraphics[width=7.0cm]{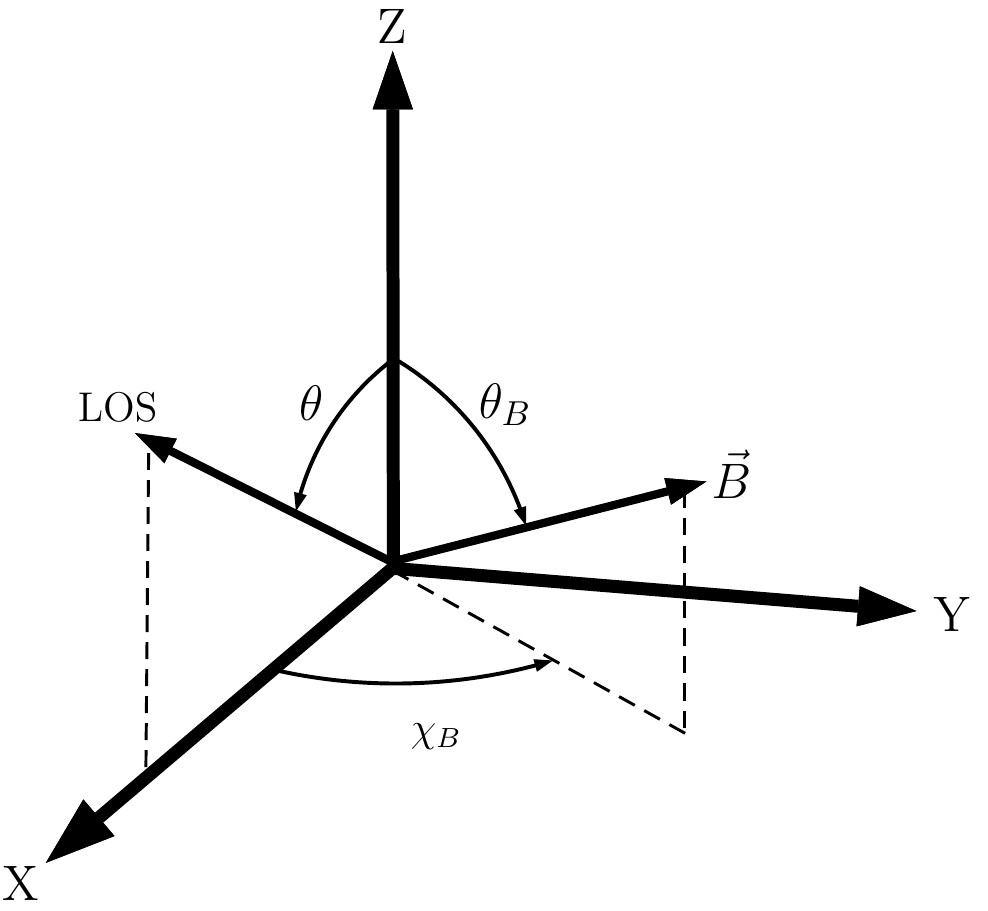}
\caption{Reference system in which the problem is formulated.}
\label{Geom}
\end{figure}
We consider a right-handed Cartesian reference system with the $Z$-axis 
directed along the local vertical and the $X$-axis directed so that the 
direction to the observer (i.e., the line-of-sight, or LOS) lies in 
the $XZ$ plane (see Fig.~\ref{Geom}). 
The LOS is thus fully determined by specifying the value of 
$\mu \equiv \cos(\theta)$, with $\theta$ the inclination with respect to the
local vertical. 
The direction of the magnetic field is specfied by its inclination ($\theta_B$) 
and azimuth ($\chi_B$), measured counter-clockwise from the $X$-axis. 
The direction for positive Stokes $Q$ is taken along the $Y$-axis, so for 
inclinations of the LOS other than $\mu = 1$ it is parallel to the limb. 

When solving the RT equation in the code described in ABT17, the inclinations
considered in the calculation are selected using a Gauss-Legendre
quadrature, with nine directions in the $\bigl[0^\circ, 90^\circ \bigr]$ interval and 
nine more in the $\bigl[90^\circ,180^\circ \bigr]$ interval. 
 A total of eight, equally spaced, azimuths are taken. 
Except where otherwise specified, the magnetic field's strength and 
orientation is assumed to be constant over all atmospheric heights.  

\section{Results for various atmospheric models}
\label{Sect:atm}
In order to provide some insight into how the Ca~{\sc i} $4227$~\AA\ spectral line is shaped when emerging from different solar regions, we compare the synthesized profiles obtained from calculations using the various semiempirical atmospheric models presented in \cite{Fontenla+93} with each other. 
We recall that model A (hereafter FAL-A) is representative of a faint region of the quiet Sun, model F (FAL-F) of a bright region of the quiet Sun, and model P (FAL-P) of a typical plage area. Model C (FAL-C) represents instead an average region of the quiet Sun. Analogous calculations have been performed using the M\textsubscript{CO} model presented in \citet{Avrett95}. This model is also representative of a region of the quiet Sun, but at heights of around $1000$~{km} its temperature and density are sensibly lower than for the previous quiet Sun models.  
Such models are static, so in this work we do not investigate the impact of atmospheric dynamics. 

\begin{figure*}[!ht]
\centering
\includegraphics[width=16.2cm]{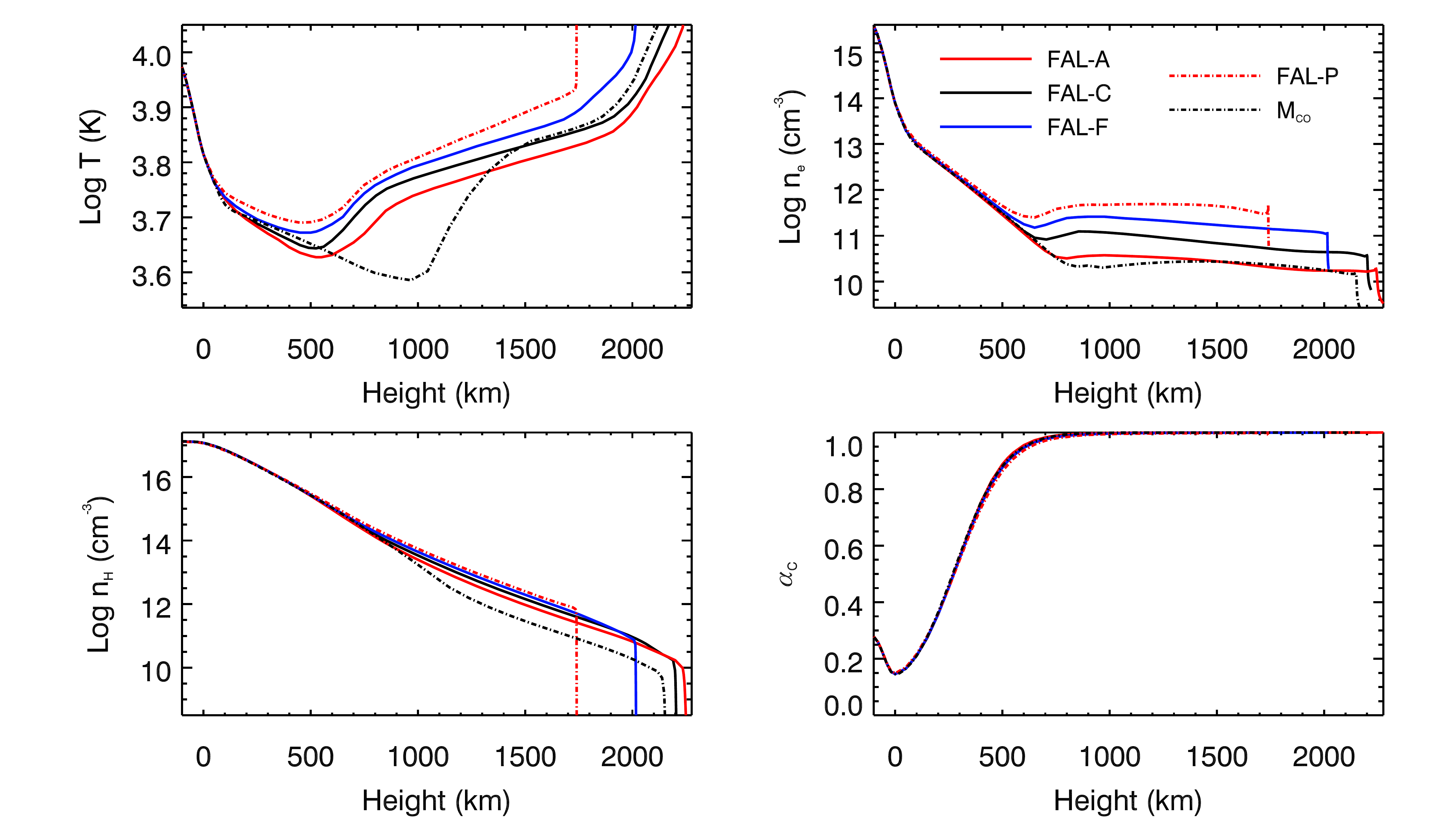}
\caption{Temperature (top left panel), electron number density (top right panel), density of neutral hydrogen atoms (bottom left panel), and coherence fraction (bottom right panel) as a function of height for the various semi-empirical atmospheric models presented in \citet{Fontenla+93} and in \citet{Avrett95}.}
\label{FAL}
\end{figure*} 
In Fig.~\ref{FAL}, the temperature $T$, the electron number density $n_e$, the number density of neutral hydrogen atoms $n_H$, and the coherence fraction $\alpha_c$ are shown as a function of height for the five atmospheric models we have considered. The coherence fraction is the probability that, after absorption, a photon will be reemitted before the atom is perturbed by an elastic collision, provided that relaxation occurs through a radiative process. Its expression can be found in ABT17. In models corresponding to more active regions, such as FAL-P, the transition region is found deeper in the atmosphere than for models corresponding to quieter regions, such as FAL-A. In the chromosphere and upper layers of the photosphere, $T$, $n_e$ and $n_H$ are all greater for models corresponding to more active regions.

\begin{figure*}[!htp]
\centering
 \includegraphics[width=16.2cm]{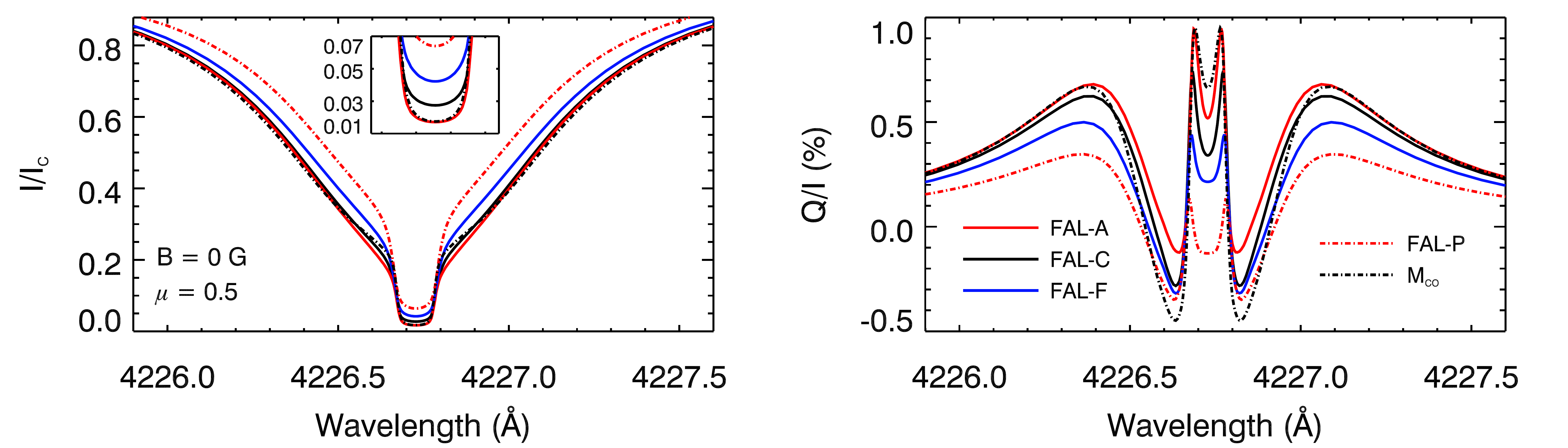}
 \caption{Emergent intensity, normalized to the continuum (left panel) and Stokes $Q/I$ (right panel) profiles for an LOS with $\mu=0.5$ calculated considering the FAL-A (red solid curves), FAL-C (black solid curves), FAL-F (blue solid curves), FAL-P (red dashed-dotted curves), and M\textsubscript{CO} (dashed-dotted black) semiempirical atmospheric models. The calculations have been performed in the absence of a magnetic field. The direction for positive Stokes $Q$ is taken parallel to the limb. The inset plot in the left panel shows the core region (between $4226.635$ and $4226.820$~\AA ) in greater detail.}
\label{FAL_mu05}
\end{figure*} 
In Fig.~\ref{FAL_mu05} the emergent intensity and $Q/I$ are calculated for an LOS with $\mu = 0.5$, in the absence of magnetic fields. Given that the positive direction for $Q$ has been taken parallel to the limb, the resulting $U$ is zero. In the absence of a magnetic field, no circular polarization is produced, and thus neither the $U/I$ nor the $V/I$ profiles are shown. 
Comparing the results for the various models, we see that the line core intensity is lowest for FAL-A, and then gradually increases for M\textsubscript{CO}, FAL-C, FAL-F, and FAL-P, for which it is {the largest}, in agreement with the findings of \citet{Supriya+14}. 

\begin{table*}
\caption{{Height and temperature at the spatial point where the optical depth is unity in the considered atmospheric models, for an LOS with $\mu = 0.5$, at the line center wavelength 
  ($\lambda_c =  4226.72$~\AA ) and at the wavelength corresponding to the maximum of $Q/I$ in the blue wing lobe {in the FAL-C model } ($\lambda_w=4226.39$~\AA ).}}
\centering
\begin{tabular}{|c||c|c||c|c|} \hline
Model & $\tau = 1$ height for $\lambda_c$ & T & $\tau = 1$ height for ${\lambda_w}$ & T \\ \hline \hline
FAL-A & $823$~{km} & $5016$~{K} & $118$~{km} & $5276$~{K} \\ \hline
FAL-C & $935$~{km} & $5819$~{K} & $116$~{km} & $5287$~{K} \\ \hline
FAL-F & $1010$~{km} & $6263$~{K} & $112$~{km} & $5366$~{K} \\ \hline
FAL-P & $1096$~{km} & $6675$~{K} & $104$~{km} & $5525$~{K} \\ \hline
M\textsubscript{CO} & $815$~{km} & $3946$~{K} & $116$~{km} & $5169$~{K} \\ \hline
\end{tabular}
\label{Table1}
\end{table*}
{For} each {of the semi-empirical models used} we have {determined} the height at which the optical depth is unity ($\tau = 1$) for an LOS with $\mu = 0.5$, both at the line center wavelength ($\lambda_c = 4226.72$~{\AA}) and at the wavelength corresponding to the $Q/I$ maximum in the blue wing lobe {for the FAL-C model} ($\lambda_w = 4226.39$~{\AA}). For each model, the temperature at such heights is shown in Table.~\ref{Table1}. It is interesting to note that, 
although the temperature at the spatial point where $\tau_{\lambda_c} = 1$ is substantially higher in the FAL-A model than in M\textsubscript{CO}, due to non-LTE effects, the line core intensity of the emergent radiation obtained in the latter model is slightly larger than in the former.

In the line wings, the intensity signals obtained in the four models presented in \citet{Fontenla+93} have the same relation to one another as in the line core, i.e., with the intensity in the FAL-A model being the lowest and the one in FAL-P being the largest. Indeed, the temperature is also lowest in FAL-A and largest in FAL-P around the wing formation region. 
However, when going from the line core into the wings, the intensity profile obtained using the M\textsubscript{CO} model presents qualitative differences with respect to the others, as a consequence of how the aforementioned atmospheric parameters are stratified in this particular model. 

As far as the linear polarization is concerned, the largest line core $Q/I$ signal is obtained in the M\textsubscript{CO} model, followed by FAL-A, FAL-C, FAL-F, and FAL-P. The $Q/I$ of the emergent radiation is an indication of the radiation anisotropy in the atmospheric regions around which the line forms. Such anisotropy is strongly influenced by the gradient of the source function, and thus the $Q/I$  is sensitive to the model's temperature gradient around the line formation region. In the wings, the relation between the $Q/I$ lobes obtained in the various models is the same as in the core, given that the temperature gradient at the heights were such lobes form is also the most pronounced in M\textsubscript{CO}, and the smoothest for FAL-P. However we point out that, when going from the core into the wings, qualitative differences are also found between the $Q/I$ obtained with M\textsubscript{CO} and the other models, for the same reason as discussed above. 

{We note that the wavelengths corresponding to the $Q/I$ wing maxima vary slightly according to the model. This is due to the fact that the position of such maxima are sensitive to the ratio of the line to the continuum opacity and the branching ratio between the} ${\mathcal R}_{\mbox{\scriptsize II}}$ and ${\mathcal R}_{\mbox{\scriptsize III}}$ {redistribution matrices, which depend on the particular srtatification of atmospheric parameters in each model. We also point out that the wavelengths at which such maxima occur depend on the LOS under consideration.} 

 Given that the polarization profiles are very sensitive to the atmospheric model (i.e, to the solar region from which the radiation originates), using the radiation polarization in order to diagnose solar magnetic fields requires that the local atmospheric parameters, such as temperature and density, be independently constrained, e.g., by means of the intensity of the emergent radiation.
 
For the sake of brevity, in the following sections we will only be considering calculations performed in the FAL-C model.
\section{The depolarizing effect of collisions}
\label{Sect:Dep}
In the absence of collisions, scattering would be perfectly coherent in the atomic rest frame and, therefore, only ${\mathcal R}_{\mbox{\scriptsize II}}$ would be required to describe it. Elastic collisions relax the frequency coherence of scattering, and make the branching ratio contained in the ${\mathcal R}_{\mbox{\scriptsize III}}$ term nonzero. {In the calculations presented in this paper, the elastic collisional rate $Q_{\mbox{\scriptsize el}}$ 
 has been taken from the RH code, which calculates it following \citet{AnsteeOmara95} and \citet{BarklemOmara97}}.
Furthermore, elastic collisions with neutral perturbers have the additional effect of relaxing atomic level polarization, i.e., of equalizing the populations of the different magnetic sublevels of a given $J$-level and relaxing quantum coherences between them, thus modifying (generally decreasing) the linear polarization fraction of scattered radiation. 
\begin{figure*}[!ht]
\centering
 \includegraphics[width=16.2cm]{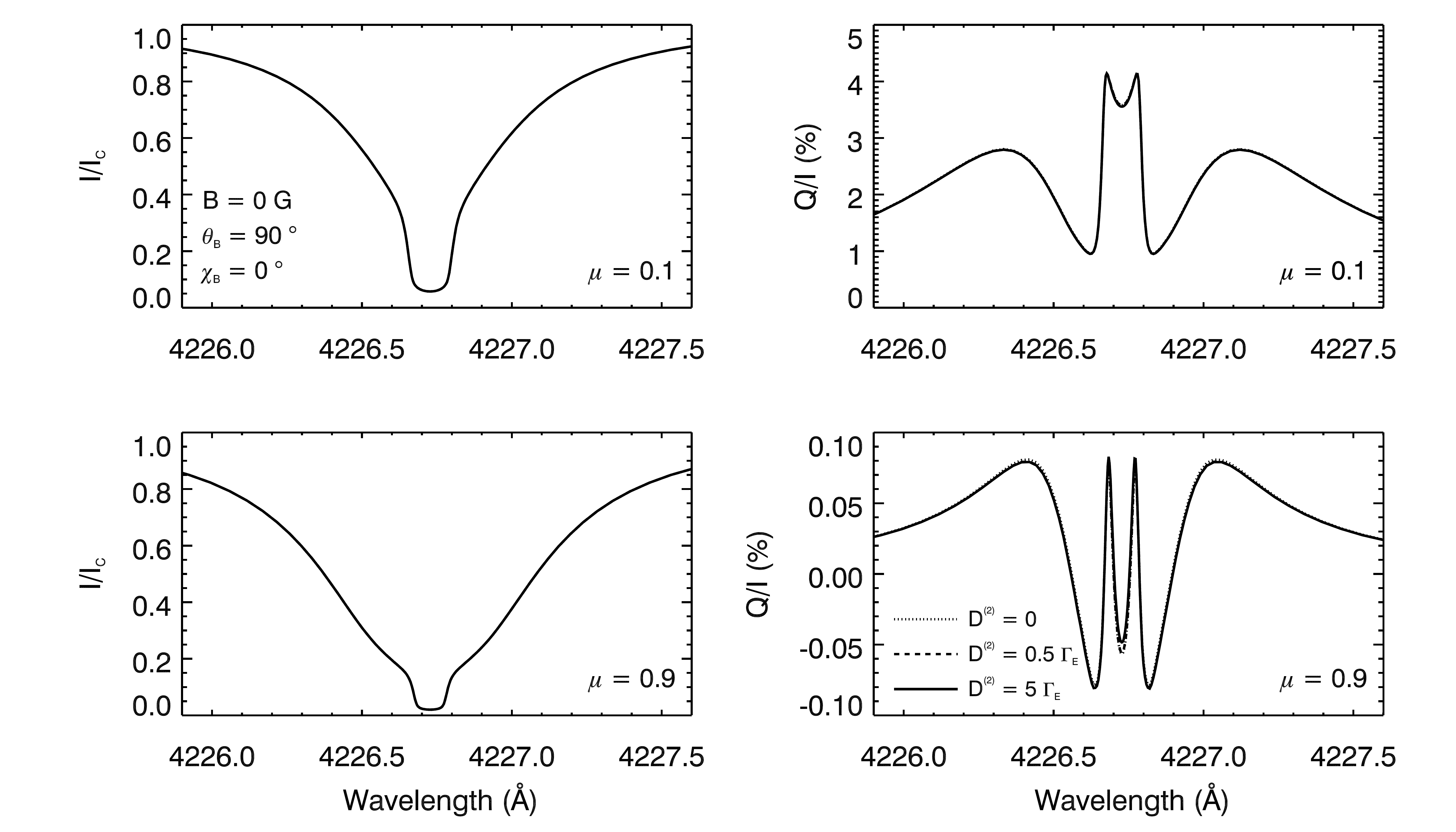}
 \caption{Emergent intensity (left column) and Stokes $Q/I$ profiles (right column) calculated considering different values of the elastic collisional depolarizing rates: $D^{(2)} = 0$ (dotted line), 
 $D^{(2)} = 0.5 \, \Gamma_E$ (dashed line), and $D^{(2)} = 5 \, \Gamma_E$ (solid line). The last case is an unrealistically large value of $D^{(2)}$, which has been considered only for illustrative purposes (see text for more details). The upper row shows the results for an LOS with $\mu = 0.1$, while the bottom row represents the same calculations for $\mu = 0.9$. No magnetic field is considered. The {reference} direction for positive Stokes $Q$ is taken parallel to the limb. Note that only the curves for $Q/I$ at $\mu = 0.9$ can be distinguished.}
\label{Fig:Depol}
\end{figure*} 
Clearly, the depolarizing effect of collisions is only contained in the ${\mathcal R}_{\mbox{\scriptsize III}}$ redistribution matrix \citep[see ABT17, and also][]{Bommier97b}. 

In the calculations presented in the previous sections, the depolarizing effect of elastic collisions has always been neglected. We now include this effect assuming that the $D^{(2)}$ multipolar component of the depolarizing rate is given by \citep[see][]{Bommier97b,Stenflo94} 
\begin{equation}
 D^{(2)} = 0.5 \, \Gamma_E , 
 \label{DepColK2}
\end{equation}  
where $\Gamma_E = Q_{\mbox{\scriptsize el}}$ is the line broadening constant due to elastic collisions. Given that the Ca~{\sc i} $4227$~\AA\ line is produced by a 0-1 transition, no $D^{(K)}$ rates with $K$ larger than 2 are involved in the calculations. The specific relation between the various $D^{(K)}$ rates depends on the type of interaction between the atom and the perturber (see LL04). For an interaction that can be described by a tensorial operator of rank 2, as is the case for a Van der Waals potential, 
\begin{equation}
\frac{D^{(1)}}{D^{(2)}} = \frac{5}{3} .
\label{DepColK12}
\end{equation} 
which coincides with the value that is obtained from a classical 
description of the atom as a collection of oscillators (see LL04). 
 We finally recall that, by definition, $D^{(0)} = 0$. 

In Fig.~\ref{Fig:Depol}, the synthesized profiles obtained when using the collisional depolarizing rates discussed above are compared to the ones obtained when they are neglected, for an LOS close to the limb ($\mu = 0.1$) and for one close to disk center ($\mu = 0.9$).  Given that the $D^{(K)}$ rates considered here are an estimate, we have also performed the calculations when multiplying such rates by a factor $10$, in order to ensure that the depolarizing effect of collisions is not being underestimated. 
 For the near-limb calculation, the depolarizing effect of collisions has no perceivable impact on the $Q/I$ profiles, either in the core of the line or in the wings, even when the depolarizing rates are artificially increased by a factor 10.

This negligible effect of depolarizing collisions could be expected, observing that the core of the Ca~{\sc i} $4227$~\AA\ line forms in the chromosphere. For an LOS with $\mu = 0.1$ 
the height corresponding to $\tau = 1$ in the core region ranges from $925$~{km} in the M\textsubscript{CO} model to $1300$~{km} in FAL-P. 
Around such heights the coherence fraction is close to one,  
and so essentially all scattering processes are described by ${\mathcal R}_{\mbox{\scriptsize II}}$ rather than by ${\mathcal R}_{\mbox{\scriptsize III}}$. Therefore, the polarization fraction of the line core is practically insensitive to depolarizing collisions.

The line wings originate much deeper in the atmosphere, where the coherence fraction is considerably smaller than unity. 
Nevertheless, 
it must be observed that the contribution to the wing emissivity brought by ${\mathcal R}_{\mbox{\scriptsize III}}$ is much smaller than the contribution brought by ${\mathcal R}_{\mbox{\scriptsize II}}$. This can be seen from the expressions of the redistribution matrices: while the emission profile of the latter is centered around the frequency of the incoming photon $\nu'$, for the former it is a generalized profile that decreases very quickly outside the Doppler core. 
In the scattering processes described by ${\mathcal R}_{\mbox{\scriptsize III}}$, the scattered photon is therefore more likely to be re-emitted in the line core region (although the frequency of the absorbed one was in the line wings), 
and so, even when the branching ratio for ${\mathcal R}_{\mbox{\scriptsize III}}$ is significant around the wing formation region, its contribution to the emissivity at these wavelengths is, comparatively, very small. Therefore, the radiation scattered at wing frequencies is largely unperturbed by elastic collisions, so their depolarizing effect is not significant.

Close to disk center, the radiation emerging both in the line core region and in the wings originates deeper in the atmosphere, and so the rate of elastic collisions is slightly larger, leading to smaller coherence fractions. Despite such differences, 
the effect of depolarizing collisions on the $Q/I$ signal is only appreciable in the line core spectral region, and only when unrealistically large $D^{(K)}$ rates are artificially imposed. We can thus conclude that the depolarizing effect of elastic collisions is completely negligible in this line, and therefore it will not be considered in the calculations presented in the rest of this paper.

\section{The impact of magneto-optical effects}
\label{Sect:MO}
In this section, we consider the impact of magnetic fields on the polarization of the emergent radiation through the anomalous dispersion terms of the propagation matrix. In the absence of lower level polarization, these terms arise because of the Zeeman splitting of the magnetic sublevels, and their effects on the radiation field are generally referred to as magneto-optical (MO) effects. 
The $\rho_V$ term is of particular interest. It couples the Stokes $Q$ and $U$ parameters when radiation travels through the magnetized plasma, thus producing a rotation of the plane of linear polarization (see LL04). 
 This effect, whose impact is proportional to the longitudinal component of the magnetic field, is also known as Faraday rotation \citep[for a more extensive discussion, see][]{Pershan67,Portis78}. As discussed in \cite{Alsina+16, Alsina+17}, this effect is significant only in the wings, where $\rho_V$, which is a linear combination of dispersion profiles and is proportional to the Zeeman splitting, can  become comparable in magnitude to $\eta_I$ already for relatively weak magnetic fields.
\begin{figure*}[htp]
\centering
\includegraphics[width=16.2cm]{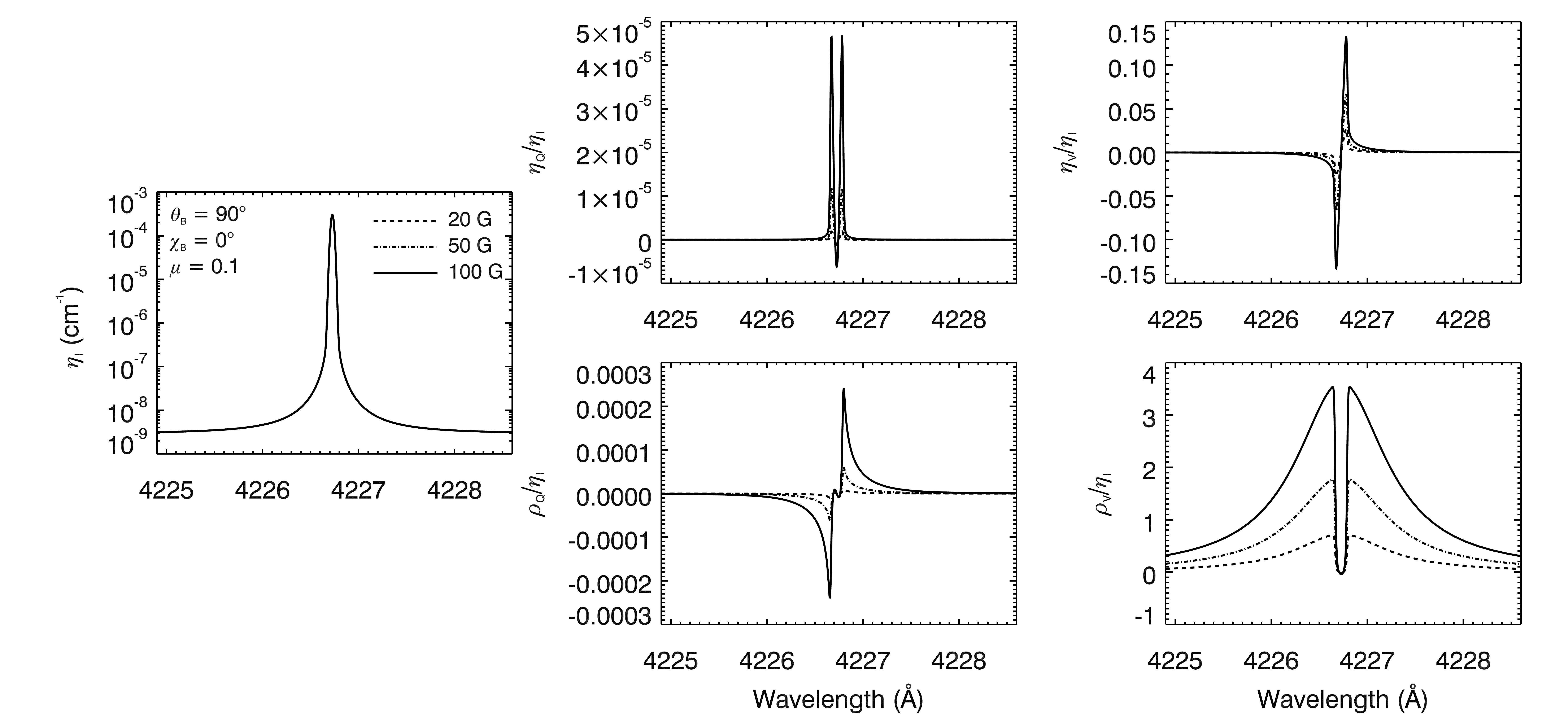}
\caption{{The variation with wavelength} of $\eta_I$ {(left panel),} $\eta_Q/\eta_I$ {(top center panel)}, $\eta_V/\eta_I$ {(top right panel)}, $\rho_Q/\eta_I$ {(bottom center panel), and} $\rho_V/\eta_I$ {(bottom right panel)}, calculated for the Ca~{\sc i} $4227$~\AA\ line at a height of $250$~km above the $\tau_{5000} = 1$ {height in} the FAL-C model, for an LOS with $\mu = 0.1$, in the presence of a horizontal magnetic field with $\chi_B = 0^\circ$ and strengths of $20$~G (dashed curves), $50$~G (dash-dotted curves), and $100$~G (solid curves). {The reference direction for positive Stokes} $Q$ {is taken parallel to the limb. The panels for $\eta_U/\eta_I$ and $\rho_U/\eta_I$ are not included because these quantities are zero for the considered geometry.}}
\label{Fig:RT}
\end{figure*} 
The spectral dependence of the $\rho_V/\eta_I$ ratio for Ca~{\sc i} $4227$~\AA\ is shown in the {bottom} right panel of Fig.~\ref{Fig:RT}, 
{taking} the atmospheric parameters from the FAL-C model, at a height of $250$~km above the $\tau_{5000} = 1$ {height}. One can clearly see that, outside the Doppler core, this ratio reaches significant values already for longitudinal fields of $20$~G, and for a $50$~G field it is {significantly} larger than unity. We note that if only the line contribution to $\eta_I$ were considered, the ratio would be constant with frequency outside the Doppler core. However, the continuum contribution to the absorption coefficient becomes increasingly dominant further into the wings. Thus, when such contribution is considered, the ratio eventually decreases to zero. {As for the other RT coefficients, we observe that the} $\eta_V/\eta_I$ {ratio is significant in the Doppler core but, as can be seen in the top right panel of Fig.~\ref{Fig:RT}, it becomes very small outside such spectral region. For the field strengths considered in the figure, which are reasonable values for quiet solar regions, both the} $\eta_Q/\eta_I$ and $\rho_Q/\eta_I$ {ratios are very small and, for our specific choice of the reference direction for positive Stokes $Q$, $\eta_U$ and $\rho_U$ are zero. As can be seen from Eqs.~\eqref{RTE} and~\eqref{Prop}, only the MO effects quantified by $\rho_Q$ and $\rho_U$ are involved in the transfer equation for Stokes $V$, and so, in the presence of the comparatively weak fields considered here, their impact on the circular polarization of the emergent radiation is inconsequential.}

We now present some applications in order to study the impact of the MO effects on the Ca~{\sc i} $4227$~\AA\ line in greater depth. We consider horizontal magnetic fields, and two different LOS, one close to the limb ($\mu = 0.1$) and one at disk center ($\mu = 1$). Since the intensity profiles are not noticeably affected for the magnetic field strengths we are considering in this work (ranging up to $100$~G), only the polarization profiles will be shown in the remainder of this section.
\subsection{Close to the limb LOS ($\mu = 0.1$)}
\label{SubSec:m01}
\begin{figure*}[htp]
\centering
\includegraphics[width=16.2cm]{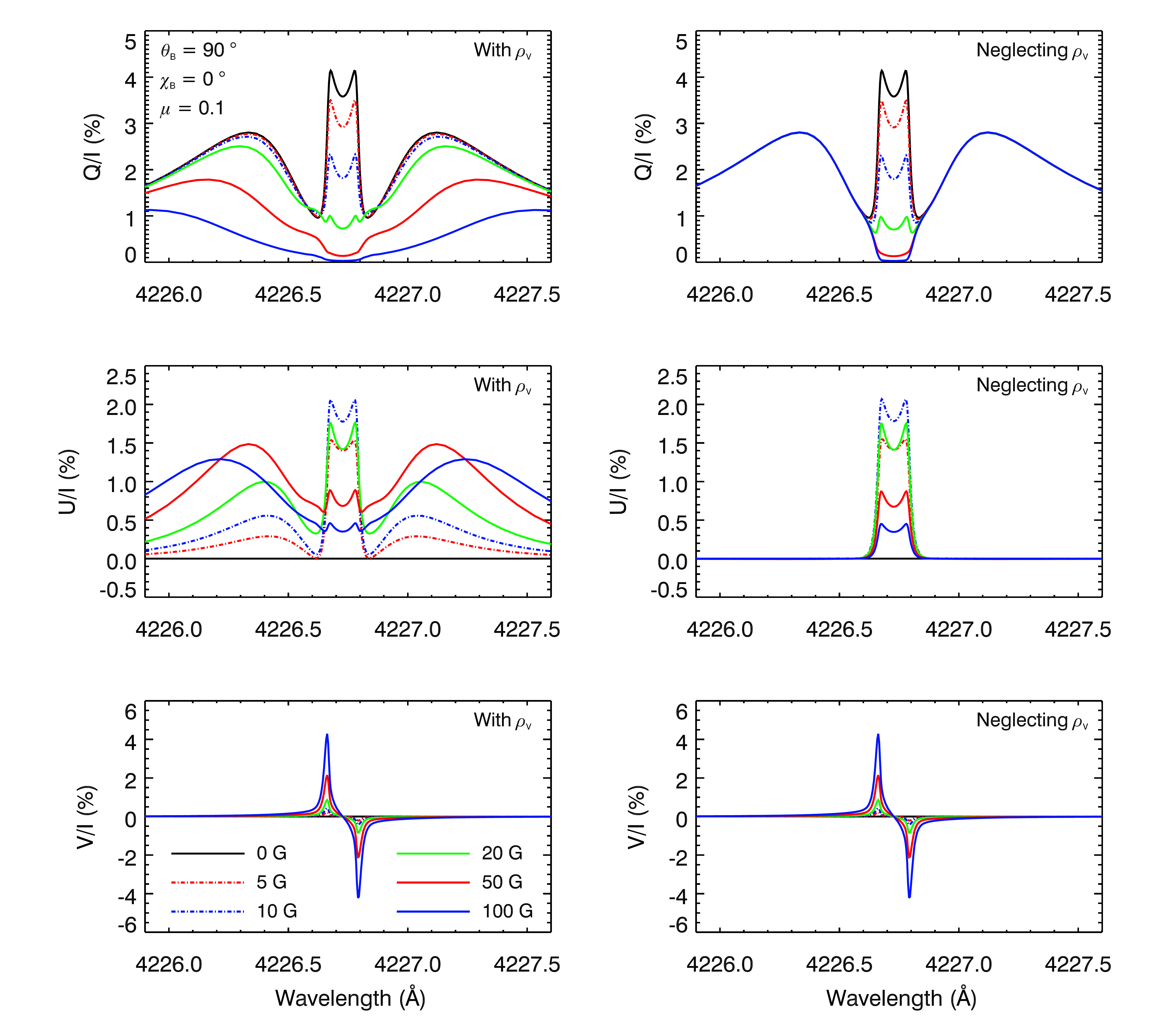}
\caption{Emergent $Q/I$ (top), $U/I$ (middle) and $V/I$ (bottom) calculated in the FAL-C semiempirical atmospheric model for an LOS with $\mu=0.1$, both considering all terms of the propagation matrix (left column) and neglecting the $\rho_V$ term when solving the radiative transfer (RT) equation (right column). The magnetic field is horizontal ($\theta_B = 90^\circ$) and contained in the plane defined by the local vertical and the LOS (i.e., $\chi_B = 0^\circ$), with various field strengths: $0$~G (black solid curve), $5$~G (red dashed-dotted curve), $10$~G (blue dashed-dotted curve), $20$~G (green solid curve), $50$~G (red solid curve), $100$~G (blue solid curve). The {reference} direction for positive $Q$ is taken parallel to the limb.}
\label{Stok_MO_900}
\end{figure*} 
In Fig.~\ref{Stok_MO_900}, near-limb calculations are shown in the presence of an almost longitudinal magnetic field, 
 both taking into account and neglecting the $\rho_V$ term in the RT equations when obtaining the self-consistent solution of the non-LTE RT problem in the atmosphere, and when calculating the emergent radiation along the considered direction. 
In the core of the line, where the $\rho_V/\eta_I$ ratio is small (unless the magnetic field is very strong), we attribute the change in $Q/I$ and $U/I$ to the Hanle effect which, for the geometry considered here, operates by reducing the degree of linear polarization of the emergent radiation and rotating its plane of polarization.

 In the wings, however, the Hanle effect is expected to have no impact, although this is strictly true only in the collisionless regime (see LL04). On the other hand, we have already pointed out that also when collisions are taken into account (and the coherence fraction can be sensibly different from unity), the emissivity in the wings is mainly due to coherent scattering. 
 These processes are, by definition, unperturbed by collisions, and therefore the scattered radiation cannot eventually be sensitive to any kind of collisionally induced wing Hanle effect.  
This can be verified both from the explicit expression of the ${\mathcal R}_{\mbox{\scriptsize II}}$ redistribution matrix (see Appendix \ref{Sect:WingHanle}), or from time-energy uncertainty arguments \citep[see][]{Faurobert92}. 
This was also confirmed by \citet{Sampoorna+09}, who modeled the Ca~{\sc i} $4227$~\AA\ line under the last scattering approximation, in the Hanle effect regime, taking the effect of collisions into account. Their calculations show that only a very small magnetic sensitivity is found in the wings when collisions are taken into account. Indeed, despite searching over a large parameter space of collisional rates and magnetic field strengths and configurations, they were unable to reproduce the wing behavior of the $Q/I$ and $U/I$ profiles observed by \citet{Bianda+03}.

 The linear polarization in the line wings is nevertheless sensitive to the magnetic field through the MO effects. As shown in Fig.~\ref{Stok_MO_900}, for fields as weak as $5$~G, significant $U/I$ signals in the wings already appear due to such effects, and for slightly stronger fields (around $10$~G), a 
 reduction in $Q/I$ is already noticeable. {Note that the wavelengths corresponding to the $Q/I$ and $U/I$ wing maxima also change when the MO effects operate, moving increasingly far away from line center as the magnetic field strength increases.} 
 
 It is {also} worth mentioning that, by taking a magnetic field whose orientation is constant over the whole atmosphere - and almost parallel to the LOS - we are considering a geometry which is especially advantageous for 
$\rho_V$ to have an impact on the linear polarization of the emergent radiation. As can be seen in the figure, the fact that such magnetic sensitivity only arises when the $\rho_V$ term is taken into account clearly demonstrates that it is not a consequence of the Hanle effect, but rather of the MO effects. {Moreover}, the circular polarization profile is obviously unaffected by the inclusion of $\rho_V$  in the calculation.

{The fact that the efficiency of the MO effects in modifying the wing polarization depends on the longitudinal component of the magnetic field can be further illustrated by considering magnetic fields with a constant inclination, and different azimuths.}  
\begin{figure*}[ht]
\centering
\includegraphics[width=13.8cm]{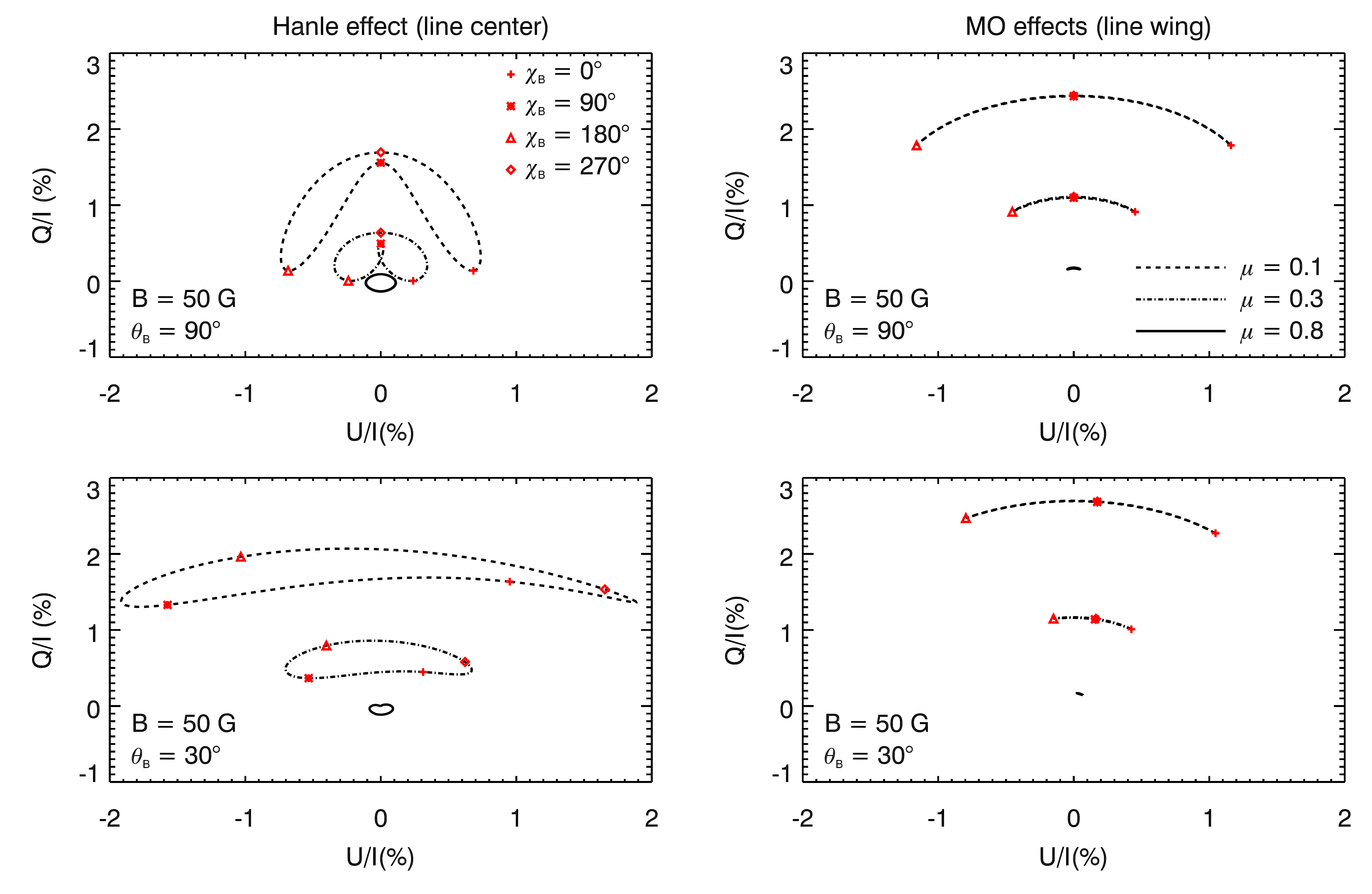}
\caption{{Hanle diagrams for the linear polarization of the emergent spectral line radiation, at line center (left column) and at the wavelength corresponding to the maximum of $Q/I$ in the blue wing (right column). Calculations have been performed in the presence of magnetic fields with $B = 50$~G, and inclinations $\theta_B = 90^\circ$ (top row) and $\theta_B = 30^\circ$ (bottom row), for a range of azimuths between $0^\circ$ and $360^\circ$, in $5^\circ$ increments. LOS with $\mu = 0.1$ (dashed curves), $\mu = 0.3$ (dashed-dotted curves), and $\mu = 0.8$ (solid curves) have been considered. The reference direction for positive Stokes $Q$ has been taken parallel to the limb.}}
\label{HanleDiag}
\end{figure*}  
{The closed curves in Fig.~\ref{HanleDiag} show the variation on the $Q/I$-$U/I$ plane (or Hanle diagrams) of the linear polarization of the emergent radiation calculated in the presence of magnetic fields with constant strength ($B = 50$~G) and inclination, while azimuths in the $\chi_B = [0^\circ,360^\circ]$ range are considered. Such diagrams are shown both {for} the line center, where the Hanle effect operates, and {for} the wavelength that corresponds to the maximum of the $Q/I$ profile 
in the blue wing, where the MO effects operate. 
Let us first consider the case of a horizontal magnetic field (see the upper panels of Fig.~\ref{HanleDiag}). 
We point out that for azimuths $\chi_B = 90^\circ$ and $\chi_B = 270^\circ$, no $U/I$ is produced in the line wing, since the magnetic field is perpendicular to the LOS. Likewise, for these azimuths no $U/I$ is produced at line center, and their $Q/I$ values differ from one another because we are considering LOS other than $\mu = 0$ or $\mu = 1$ \citep[e.g.,][]{TrujilloBueno01}. 
Moreover, for horizontal fields the Hanle diagrams are symmetric with respect to $U/I = 0$, both for line center and for the line wing maxima. This is because the rotation of the plane of linear polarization is produced, both for the Hanle effect and for the MO effects, by the component of the magnetic field that is parallel to the LOS. For a horizontal magnetic field with azimuth $\chi_B$, its component along the LOS is opposite to that of a magnetic field with $\chi_B + 180^\circ$. As a consequence, for radiation reaching an observer from a spatially unresolved region in which various horizontal magnetic fields are present, such that their azimuths are equally distributed in any direction, the overall U/I signal will be canceled out, for any LOS, both at line center and in the wings. 
The situation is clearly different if a magnetic field with a different inclination is considered, as is illustrated for $\theta_B = 30^\circ$ in the lower panels of Fig.~\ref{HanleDiag}. In this case, the Hanle diagrams are not symmetric around $U/I = 0$, since the longitudinal components of the magnetic fields are not generally symmetrically distributed around the LOS. 
Thus, if the radiation reaches the observer from a spatially unresolved region with magnetic fields of inclination $30^\circ$ and equally distributed azimuths, there will be a remaining non-zero $U/I$ signal.}

It is also interesting to study the variation of the total linear polarization fraction 
\begin{equation}
P_L = \sqrt{\frac{Q^2 + U^2}{I^2}} \, ,
\label{DefPolFrac}
\end{equation}
and the linear polarization angle 
\begin{equation}
\alpha = \frac{1}{2} \tan^{-1}\biggl(\frac{U}{Q} \biggr) \, ,
\label{PolAng}
\end{equation}
 induced by such MO effects.

\begin{figure*}[!htp]
\centering 
\includegraphics[width=16.2cm]{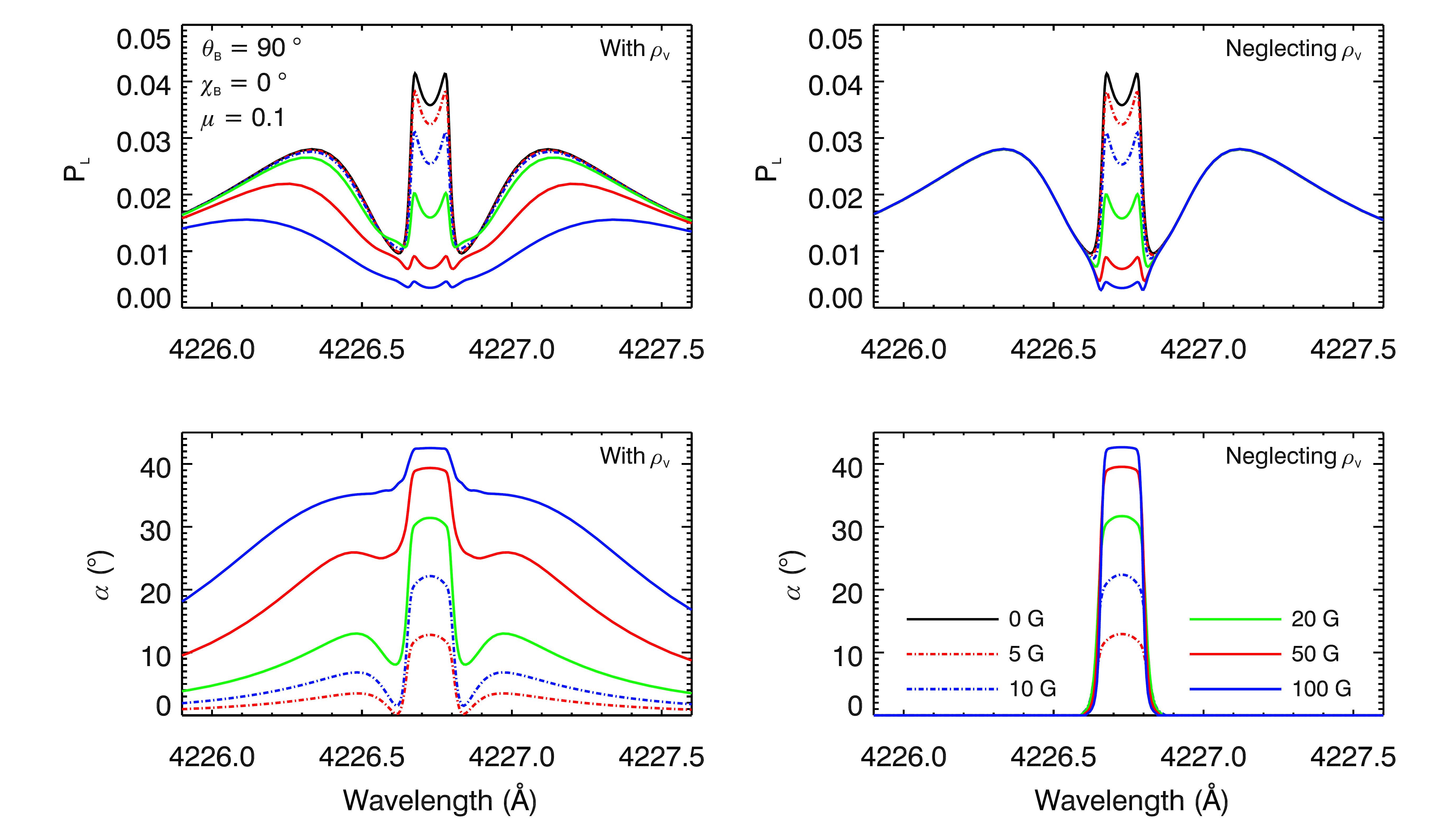}
\caption{Total linear polarization fraction (top row) and polarization angle (bottom row) 
of the emergent radiation calculated by considering the full propagation matrix in the RT equations (left column), and by neglecting the $\rho_V$ coefficient (right column). The same LOS and magnetic field orientation and strengths as in Fig.~\ref{Stok_MO_900} have been considered.}
\label{Stok_MO_LIN_900}
\end{figure*} 
In Fig.~\ref{Stok_MO_LIN_900}, $P_L$ and $\alpha$ are shown for the same geometry as in {Fig.~\ref{Stok_MO_900}}. In the core of the line one encounters the typical behavior of the Hanle effect, i.e., the linear  polarization fraction is reduced as the magnetic field increases, and $\alpha$ increases as the Hanle {effect} gives rise to a $U/I$ signal. When Hanle saturation is approached, further increases in the field strength produce smaller changes in the linear polarization fraction and angle. 
 When all terms are included in the propagation matrix, the behavior in the wings (due to $\rho_V$) and in the core (due to the Hanle effect) is qualitatively similar, but with some differences {(see also Fig.~\ref{HanleDiag})}. For weak fields (around $5$~G), the change in the linear polarization fraction is much greater in the core than in the wings, but as the magnetic field becomes larger and Hanle saturation is approached, the change with magnetic field of the linear polarization fraction in the wings begins to be greater than in the core. Also, as the magnetic field increases, both the reduction of the linear polarization fraction and the increase of the linear polarization angle extend increasingly further into the wings, as $\rho_V$ becomes significant compared to $\eta_I$ (which includes the continuum contribution). 

\textit{A priori}, one may not have expected Faraday rotation to cause a reduction of the {total} linear polarization fraction, as is found in the wings. Indeed, as is demonstrated in Appendix~\ref{App:Far}, if the only nonzero coefficients of the propagation matrix in a given spatial region are $\eta_I$ and $\rho_V$, and the region under consideration is non-emitting, then the linear polarization fraction of the radiation crossing this region remains constant, although the linear polarization angle will be modified. However, if polarized radiation is emitted within this region, and since the rotation of the plane of polarization induced by $\rho_V$ depends on the distance traveled through the magnetized material, the polarization angle of radiation originating at different points along the ray path is changed by different amounts. As a result, the total polarization fraction may effectively be reduced. 

In order to reinforce our statement that the magnetic field dependence at wing and core frequencies is caused by different effects, in
\begin{figure*}[hb]
\centering
\includegraphics[width=13.8cm]{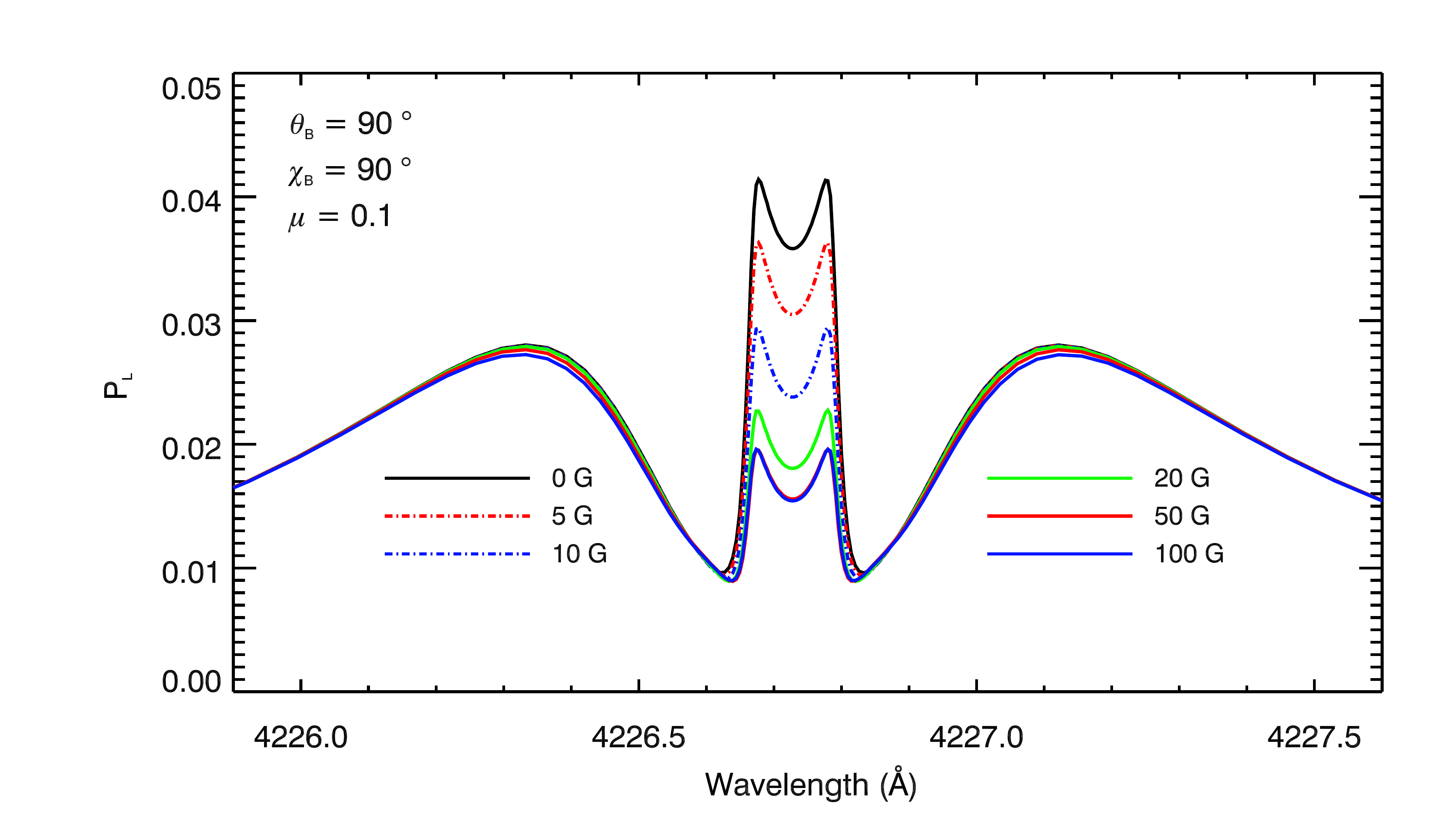}
\caption{Linear polarization fraction of the emergent radiation, calculated for a LOS with $\mu=0.1$, in the presence of a horizontal magnetic field with $\chi_B = 90^\circ$ (i.e., perpendicular to the LOS). The colored curves indicate the same field strengths as in the two previous figures.}
\label{Stok_LIN_9090}
\end{figure*}  
Fig.~\ref{Stok_LIN_9090} we show $P_L$ for radiation emerging at $\mu = 0.1$ in the presence of a horizontal field with $\chi_B = 90^\circ$, i.e., perpendicular to the LOS, accounting for Zeeman
splitting and for the full propagation matrix. In the core, for the considered magnetic field geometry, the Hanle effect produces a reduction of the emergent polarization with no change in the linear polarization angle. In the wings, no rotation of the plane of linear polarization is produced by MO effects, because the magnetic field has no component along the LOS.

The linear polarization fraction, however, shows a small magnetic sensitivity in the wings, and we have checked that this only happens when $\rho_V$ is accounted for.
This is a consequence of the fact that the polarization properties of the emissivity at a given spatial point depend, through the redistribution matrix, on those of the radiation reaching such point from all directions. For some of these directions - those for which the magnetic field has a longitudinal component - 
 $\rho_V$ is nonzero, and it actually modifies the polarization of the pumping radiation field. As a result, the emitted radiation is somewhat depolarized. This mechanism was already identified and discussed in \citet{Alsina+16}, and it will be further addressed in the next section.

{In the calculations presented in this work, the linear polarization of the continuum has always been included. This polarization is produced by Thomson and Rayleigh scattering, and it is parallel to the limb\footnote{{Taking the reference direction for positive Stokes $Q$ parallel to the limb, the continuum polarization produced by Thomson and Rayleigh scattering is taken into account through a continuum emissivity term $\varepsilon^c_Q$.}}. }
\begin{figure*}[ht]
\centering
\includegraphics[width=13.8cm]{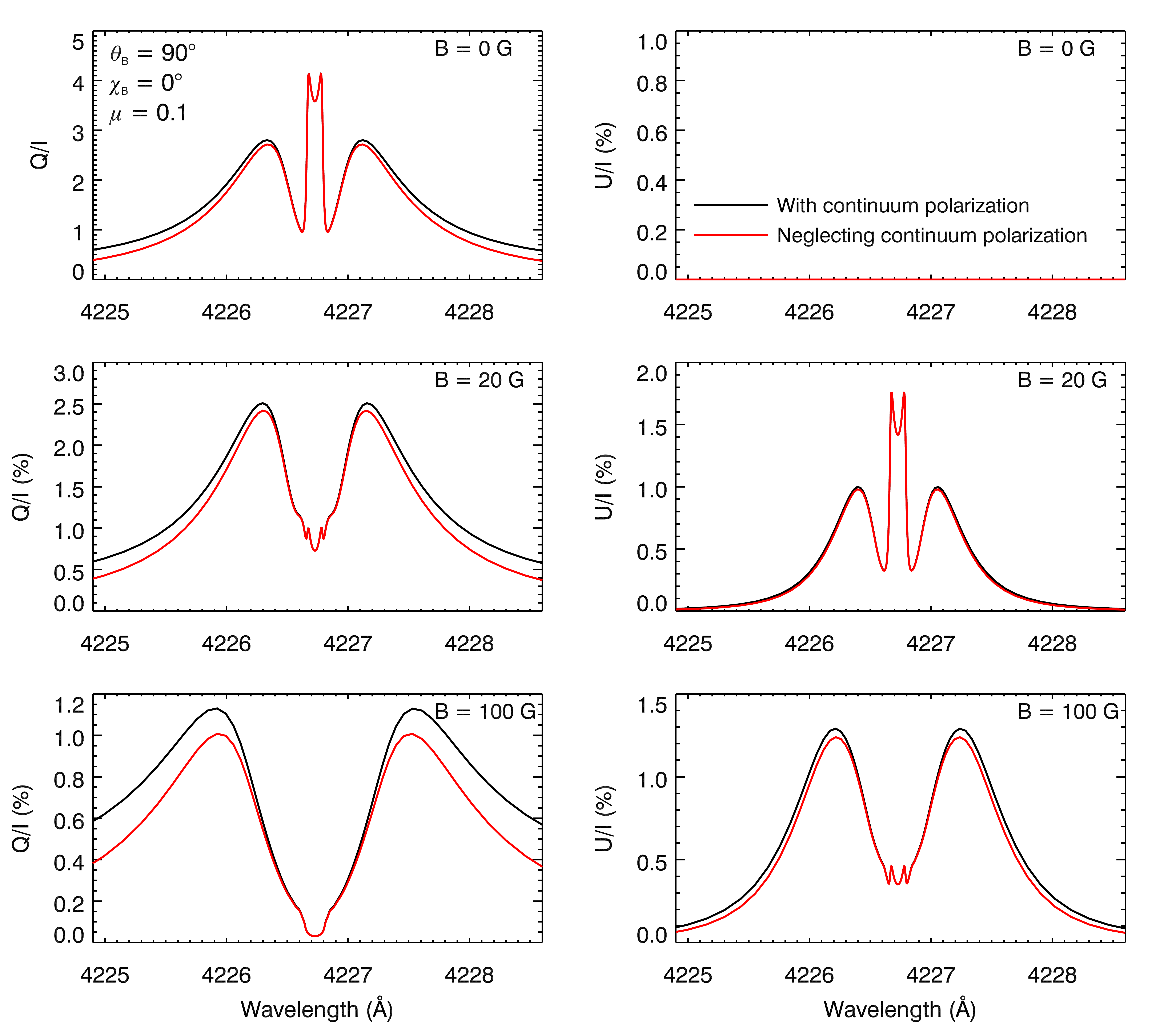}
\caption{{Emergent $Q/I$ (left column) and $U/I$ (right column) calculated in the FAL-C semi-empirical model for an LOS with $\mu = 0.1$, obtained both taking into account (black curves) and neglecting (red curves) the polarization of the continuum radiation. Calculations have been performed in the absence of a magnetic field (top row) and in the presence of a magnetic field with inclination $\theta_B = 90^\circ$ and azimuth $\chi_B = 0^\circ$ with strengths of $20$~G (middle row) and $100$~G (bottom row). The reference direction for positive Stokes $Q$ has been taken parallel to the limb.}}
\label{MO_cont}
\end{figure*}  
{The impact of the continuum polarization on the emergent Stokes profiles can be seen in Fig.~\ref{MO_cont}, where the $Q/I$ and $U/I$ profiles emerging at $\mu = 0.1$, calculated taking into account and neglecting continuum polarization, are compared. We show the profiles obtained in the absence of a magnetic field (upper panels), as well as in the presence of horizontal magnetic fields with azimuth $\chi_B = 0^\circ$, and field strengths of $20$~G (middle panels) and $100$~G (bottom panels). As expected, the continuum polarization has no influence at all in the line core region, where the line opacity is much larger than that of the continuum. Its contribution is, on the other hand, clearly appreciable in the wings of the $Q/I$ and $U/I$ profiles, becoming more significant as the magnetic field increases. 
The sensitivity of $U/I$ to the continuum is particularly interesting, because the continuum polarization is parallel to the limb, and thus it should only contribute to Stokes $Q$. This sensitivity of $Q/I$ and $U/I$ to the continuum polarization is in fact another signature of the MO effects. Indeed, linearly polarized photons, produced by Rayleigh and Thomson scattering processes, are present deep in the atmosphere. These photons subsequently interact with atoms, being thereby subject to the MO effects, which rotate their plane of linear polarization, thus modifying the emergent Stokes $Q/I$ and $U/I$ signals with respect to the case in which the continuum polarization is not taken into account. }  
\subsection{{Disk center LOS} ($\mu = 1$)}
\label{SubSect:m1}
Now we consider an LOS with $\mu = 1$, i.e., the forward scattering case. 
\begin{figure*}[htp]
\centering
\includegraphics[width=16.2cm]{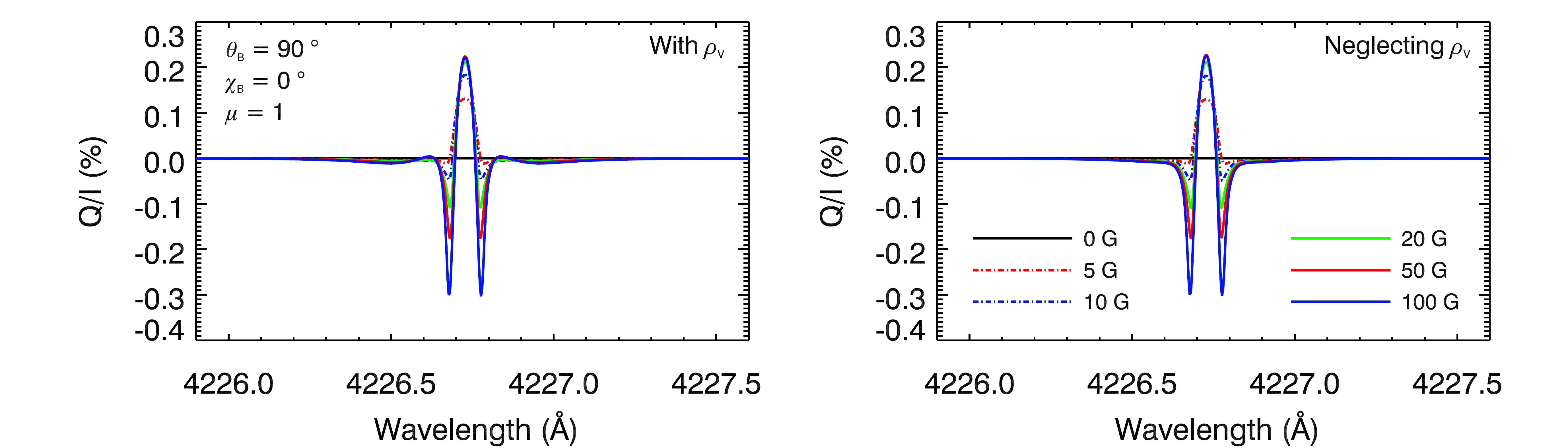}
\caption{Emergent Stokes $Q/I$ profiles calculated for an LOS with $\mu=1$, both considering all terms of the propagation matrix (left panel) and neglecting the $\rho_V$ term when {solving} the RT equation (right panel). A horizontal ($\theta_B = 90^\circ$) magnetic field with azimuth $\chi_B = 0^\circ$ has been considered, and the colored curves represent the same field strengths as in {Fig.~\ref{Stok_MO_900}}. The {reference} direction for positive $Q$ is taken perpendicular to the magnetic field.}
\label{Stok_Forward_900}
\end{figure*}  
In Fig.~\ref{Stok_Forward_900}, forward scattering polarization profiles are shown in the presence of horizontal fields of various strengths. The direction for positive $Q$ is taken perpendicular to the magnetic field. Stokes $U$ and $V$ are zero in this case, and therefore they are not shown. At line center a $Q/I$ signal arises in the presence of a horizontal magnetic field because of the Hanle effect in forward scattering \citep[e.g.,][]{TrujilloBueno01}. 
It is clear that the negative dips seen next to the line center, whose amplitudes scale with the square of the magnetic field strength, are caused by the Zeeman effect in emission.

Further into the wings, a slight sensitivity to the MO effects described by $\rho_V$ is found. 
Given that the magnetic field is perpendicular to this LOS, $\rho_V$ is zero for this direction. However, the radiation emerging with LOS $\mu = 1$ is affected by the MO effects through the same mechanism discussed in the previous section \citep[see][for more details]{Alsina+16}, which is related to the nonlocal nature of radiative transfer. The radiation propagating in directions upon which the magnetic field vector has a nonzero projection experiences a rotation of its plane of linear polarization, in spectral and spatial regions where $\rho_V/\eta_I$ is significant. This causes a breaking of the axial symmetry of the radiation field in the wings, which is at the origin of the small magnetic sensitivity that is found at such wavelengths. 

It is interesting to observe that the wing signals found in the Ca~{\sc i} $4227$~\AA\ line due to the aforementioned mechanism are much weaker than the ones with the same origin obtained in the Mg~{\sc ii} $k$ line \citep[see][]{Alsina+16}. The reason for this is related to the fact that the wings of the Ca~{\sc i} $4227$~\AA\ line form much lower in the atmosphere, at the photospheric level. The linear polarization fraction of the radiation propagating at such atmospheric heights, in any direction, is very small. Therefore, the MO effects, which rotate its plane of linear polarization, of course have very little impact on it. 
As a result, the impact of the MO effects on the linear polarization signals at $\mu = 1$ is very small, even for magnetic field strengths of $100$~G. Thus, disentangling the magnetic origin of the wing linear polarization signals from the atmosphere's thermodynamical properties would be a very challenging task, even without considering the spectral smearing of the measured signals due to the finite resolution of any instrument.

 We have also emulated such spectral smearing by convolving the synthesized Stokes profiles with a Gaussian function. 
In particular, we have observed that the dip shown by the $Q/I$ profile in the line-core region, which becomes more pronounced going from the limb to the disk center, can no longer be distinguished at $\mu=0.1$ when the FWHM is larger than $50$~{m\AA}, while for an LOS with $\mu=0.9$, it is lost if the FWHM is 65~m{\AA}, or larger. 

\subsection{Comparison with observations}
\label{SubSect:Diag}
\cite{Bianda+03} presented near-limb observations of the Ca~{\sc i} $4227$~\AA\ line, in which both a significant spatial variation in $Q/I$ and significant $U/I$ signals were found at wing wavelengths. Given that the Hanle effect is restricted to the line core region, both these authors and others \citep[see][]{Sampoorna+09} suggested a non-magnetic origin for this behavior, 
such as the symmetry breaking produced by the presence of local atmospheric inhomogeneities. 
While we cannot exclude that such inhomogeneities also play a role, our key point is that the MO effects introduce $U/I$ wing signals and a clear magnetic sensitivity in the wings of $Q/I$ and $U/I$, which  imply spatial variability given that the 
magnetic field of the solar atmosphere is not spatially homogeneous. 

The presence of a magnetic field in the formation region of the wings can, via a modification of the polarization angle induced by MO effects, produce a magnetic sensitivity in the wings.
Clearly, the strength and orientation of the magnetic field at the formation region of the wings is not necessarily the same as that of the field around the formation region of the core. As shown below, this could explain why $U/I$ at the core and in the wings can have different signs, as is found in Fig.~3 of \cite{Bianda+03} for an observation at $\mu = 0.3$. 

\begin{figure*}[!htp]
\centering 
\includegraphics[width=16.2cm]{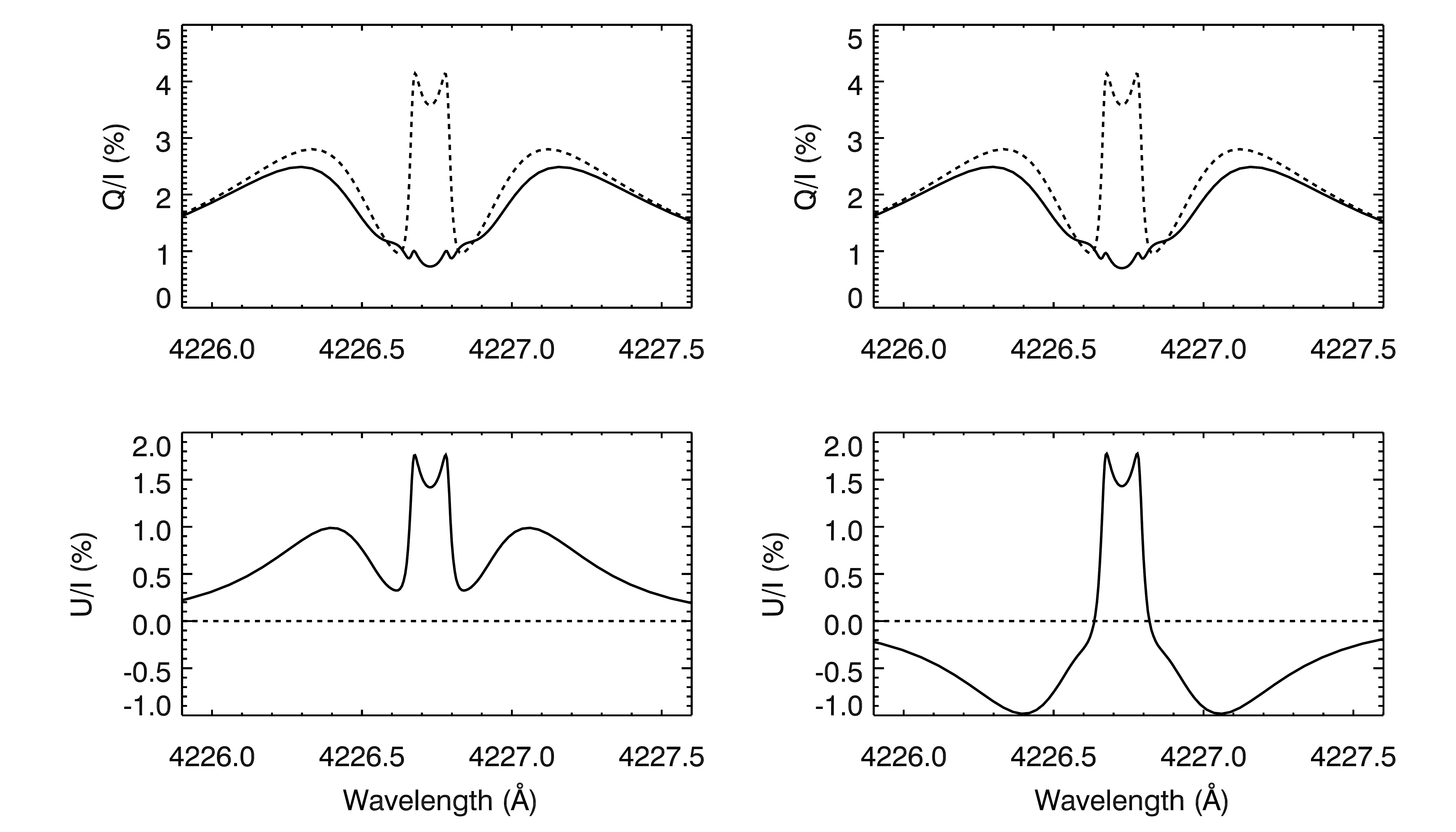}
\caption{Emergent Stokes $Q/I$ (top row) and $U/I$ (bottom row) calculated for an LOS with $\mu = 0.1$. The solid curves correspond to calculations in the presence of $20$~G horizontal magnetic fields. In the left column, the magnetic field is oriented with an azimuth of $\chi_B = 0^\circ$ at all spatial points, while in the right column its azimuth is $\chi_B = 0^\circ$ at heights of $700$~km and above, and $\chi_B = 180^\circ$ below this height. The dashed curves correspond to calculations in the absence of a magnetic field. The {reference} direction for positive Stokes $Q$ is taken parallel to the limb.}
\label{Stok_Ca_Inv}
\end{figure*} 
In Fig.~\ref{Stok_Ca_Inv} this is illustrated through calculations of the emergent linear polarization profiles for an LOS with $\mu = 0.1$.  The results obtained in the case in which a $20$~G horizontal magnetic field with azimuth $\chi_B = 0^\circ$ is present at all spatial points are compared to the case in which the orientation of the magnetic field changes at a height of $700$~km above the photospheric surface, which is below the formation height for the core, but above the height where the $Q/I$ wing lobes form. At $700$~km and above, we consider a $20$~G horizontal magnetic field with azimuth $\chi_B = 0^\circ$, while below this height the orientation of the magnetic field is inverted, setting $\chi_B = 180^\circ$. 

As expected, both the emergent $Q/I$ and $U/I$ profiles coincide in the core region, but this is not the case in the wings. 
Comparing with the profiles obtained in the absence of a magnetic field, it can be clearly seen that the MO effects cause a decrease of $Q/I$, no matter the sign of $\rho_V$ (which depends on the orientation of the magnetic field relative to the LOS). 
 A $U/I$ signal also appears, and it is positive or negative, depending on the sign of $\rho_V$. This can be easily understood by observing that, in the absence of magnetic fields, the linear polarization plane is parallel to the limb (so that $U/I=0$ with our definition of the reference direction), and that $\rho_V$ produces a rotation of the plane of linear polarization, which can be clockwise or counter-clockwise, depending on its sign.

In conclusion, if we assume that the magnetic field component along the LOS has opposite directions in the formation regions of the core and wings, then we can get a $U/I$ signal with opposite signs in the core and wings. Such signals have indeed been found in the aforementioned observations of \citet{Bianda+03}.

Accounting for MO effects in lines such as Ca~{\sc i} $4227$~\AA\ enhances their diagnostic capabilities, since the core and the wing of the line give information on the thermal and magnetic properties of the atmosphere over a large range of atmospheric heights. It is interesting to note that, via such MO effects, the wing polarization is sensitive to magnetic fields with strengths of the order of the Hanle critical field. This contrasts with the linear polarization due to the Zeeman effect in emission, which requires stronger fields in order to produce observable signals.

 The circular polarization produced by the Zeeman effect is significant already at field strengths in the vicinity of the Hanle critical field, although its signals are found mainly at frequencies closer to the core and, therefore, it is not as suitable for obtaining information on the magnetism at lower atmospheric heights.
{It is well-known that the signatures of the Zeeman effect in emission tend to vanish, due to cancellation effects, in the presence of magnetic fields that are tangled at sub-resolution scales (e.g., in the presence of an isotropically distributed magnetic field). Concerning the signatures of the MO effects in the wings of the linear polarization profiles, we note that the $U/I$ wing signals may also be canceled in the presence of similarly unresolved magnetic fields, although a net reduction in $Q/I$ may still be appreciable if the magnetic field is structured at scales larger than the mean free path of the line's photons (see Fig.~\ref{Stok_Ca_Inv}). On the other hand, in the presence of magnetic fields with mixed polarities at smaller scales (e.g., an isotropic microturbulent field), the field-averaged $\rho_V$ will be zero \citep[see Eq.~48 of][]{Alsina+17}}, and thus the MO effects will have no observable impact on the linear polarization signals in the wings. 
{This contrasts with the signatures of the Hanle effect, which reduce the overall polarization fraction of scattered radiation even in the presence of such fields
 \citep[see the appendix of][]{TrujilloBuenoMansoSainz99}.}
 
\section{Concluding comments}
\label{Sect:Concl}
A few years ago, we theoretically discovered that, in strong resonance lines for which the effects of PRD produce broad $Q/I$ profiles with sizeable amplitudes in their wings, the magneto-optical $\rho_V \, U$ and $\rho_V \, Q$ terms of the transfer equations for Stokes $Q$ and $U$, respectively, produce broad $U/I$ profiles with large wing signals, as well as an interesting magnetic sensitivity {in the wings} of both $Q/I$ and $U/I$. That finding, summarized in Sect.~\ref{Sect:Introd}, came after having developed a complex radiative transfer code for investigating the intensity and polarization of resonance lines, taking into account PRD phenomena and the joint action of scattering processes and the Hanle and Zeeman effects for arbitrary magnetic fields. 

Motivated by the {enigmatic} spectropolarimetric observations of the Ca~{\sc i} $4227$~\AA\ resonance line by \citet{Bianda+03}, which showed broad $Q/I$ and $U/I$ profiles and spatial variability in their wings, we have applied the radiative transfer code described in ABT17 and we have carried out a detailed radiative transfer investigation using semiempirical models of the solar atmosphere, taking into account the above-mentioned physical mechanisms. 

The main result of our investigation of the Ca~{\sc i} $4227$~\AA\ resonance line is that the joint action of PRD and MO effects produces broad linear polarization profiles with large amplitudes in their  wings, with a clear magnetic sensitivity to field strengths as low as $5$~G in the wings of both $Q/I$ and $U/I$. We consider the illustrative examples of Fig.~\ref{Stok_Ca_Inv}, which are qualitatively similar to the observed profiles shown in Fig.~3 of \citet{Bianda+03}, as especially remarkable. 

It is therefore very important to note that the magnetic sensitivity of the linear and circular polarization in the Ca~{\sc i} $4227$~\AA\ line is controlled by the following mechanisms: (a) the familiar Zeeman effect, which produces measurable circular polarization signals when having {spatially resolved} fields with strengths larger than $10$~G in the upper solar photosphere, (b) the Hanle effect in the core of the $Q/I$ and $U/I$ profiles, which is sensitive to magnetic fields as low as $5$~G in the lower solar chromosphere, and (c) the joint action of PRD and MO effects, which creates $U/I$ wing signals and makes the wings of $Q/I$ and $U/I$ sensitive to similarly weak magnetic fields in the solar photosphere. We point out that the observable $U/I$ {wing} signals produced by the MO effects may be masked if magnetic fields with mixed polarities are present in spatially unresolved regions of the solar photosphere, {although signatures of such MO effects will still be appreciable in $Q/I$ if the magnetic field is structured at scales larger than the mean free path of the line's photons}. Our additional finding that the linear polarization signals are practically insensitive to the depolarizing effect of elastic collisions will facilitate the development of suitable techniques for exploring the photosphere and chromosphere of the Sun via spectropolarimetry in the Ca~{\sc i} $4227$~\AA\ line. 

{Moreover, we note that the emergent radiation may be sensitive to $\rho_V$ even when the magnetic field is perpendicular to the LOS. As shown in \citet{Alsina+16}, this kind of effect is expected to be very significant in the wings of the Mg~{\sc ii} $k$ line, while the calculations performed in this work, considering a LOS with $\mu=1$ and a horizontal magnetic field, reveal that it is much less evident in the Ca~{\sc i} $4227$~{\AA} line. We attribute this lack of sensitivity to MO effects to the fact that the wings of the Ca~{\sc i} $4227$~\AA\ line form in the photosphere. In such atmospheric regions, the linear polarization degree of the radiation propagating in directions parallel to the magnetic field is very small, and as a result the MO effects only slightly modify the radiation field. As a consequence, the radiation scattered along the local vertical by such radiation field is likewise practically unaffected by the MO effects. }
 
{Finally}, we emphasize that the MO effects we have discussed here produce observable signatures in the $Q/I$ and $U/I$ wings of strong resonance lines for which PRD phenomena are important, and that such spectral lines are typically found in the blue and ultraviolet regions of the solar spectrum (e.g., Ca~{\sc i} $4227$~\AA , Sr~{\sc ii} $4078$~\AA , Ca~{\sc ii} $H$ \& $K$, Mg~{\sc ii} $h$ \& $k$, etc.). In our opinion, this is an extra scientific reason to develop ground, balloon and space telescopes with the capability of doing also high-precision spectropolarimetry in such relatively unexplored spectral regions.    

\acknowledgements
{We thank the anonymous referee for useful comments that motivated us to clarify the impact of the polarization of the continuum radiation on the MO effects.} 
Financial support by the Spanish Ministry of Economy and Competitiveness through projects \mbox{AYA2014-60476-P} and \mbox{AYA2014-55078-P} is gratefully acknowledged. JTB wishes to acknowledge also the funding received from the European Research Council (ERC) under the European Union's Horizon 2020 research and innovation programme (grant agreement No 742265). EAB is grateful to the Fundaci\'on La Caixa for financing his Ph.D. grant. 

\appendix
\section{The change in polarization fraction due to Faraday rotation}
\label{App:Far}
One could expect that $\rho_V$, which couples Stokes $Q$ and $U$ in the transfer equation, should preserve the linear polarization fraction of the radiation propagating in the medium. In order to study this in greater depth, we consider the evolution operator formalism \citep[e.g.,][]{LandiLandi85}. 
 By definition, the evolution operator $\hat{O}(s,s')$, is a $4 \times 4$ matrix which, when applied to the Stokes vector at point $s'$, yields the Stokes vector at point $s$ (with $s \geq s'$)
\begin{equation}
\vec{I}(s) = \hat{O}(s,s') \vec{I}(s') \, .
\label{EvopDef}
\end{equation}
As can be found from Sect.~8.3 of LL04, in the case in which the only nonzero elements of the propagation matrix are $\eta_I$ and $\rho_V$, the evolution operator takes the form
\begin{equation}
\hat{O}(s,s') = \mathrm{e}^{- H_I} \left(\begin{array}{c c c c}
1 & 0 & 0 & 0 \\
0 & \cos R_V & -\sin R_V & 0 \\
0 & \sin R_V & \cos R_V & 0 \\
0 & 0 & 0 & 1 \end{array} \right) \, ,
\label{EvEtIRhoV}
\end{equation}
where 
\begin{equation}
H_I = \int_{s'}^{s} \! \eta_I(s'') \, \mathrm{d}s'' \, , \quad R_V = \int_{s'}^{s} \! \rho_V(s'') \, \mathrm{d}s''
\label{ParamsEtIRhoV}
\end{equation}
So the Stokes parameters, after applying the evolution operator, become
\begin{subequations}
\begin{align}
I(s) & = \mathrm{e}^{- H_I} I(s') \, , \quad V(s) = \mathrm{e}^{- H_I} V(s') \\
Q(s) & = \mathrm{e}^{- H_I} \bigl[Q(s') \cos R_V  - U(s') \sin R_V \bigr] \\
U(s) & = \mathrm{e}^{- H_I} \bigl[Q(s') \sin R_V + U(s') \cos R_V \bigr] \, .
\end{align}
\label{StokesOP}
\end{subequations}
and together with Eqs.~\eqref{StokesOP} it can easily be found that 
\begin{equation}
\frac{Q(s)^2 + U(s)^2}{I(s)^2} = \frac{Q(s')^2 + U(s')^2}{I(s')^2} \, ,
\label{PL_Noemit}
\end{equation}
so the polarization fraction of the radiation does not change when it crosses this medium, independently of the values of $\eta_I$ and $\rho_V$. However, this situation changes when we account for the fact that the material between points $s'$ and $s$ may also emit radiation. In this case, the Stokes vector at point $s$ can be obtained through
\begin{equation}
\vec{I}(s) = \int_{s'}^{s} \! \hat{O}(s,s'') \, \vec{\varepsilon}(s'') \, \mathrm{d}s'' + \hat{O}(s,s') \vec{I}(s') \, .
\label{EvopRTE}
\end{equation}
For simplicity we consider the case in which the Stokes parameters at point $s'$ are all zero, so only the polarization fraction for the radiation emitted in the slab between $s'$ and $s$ will be considered; i.e., we disregard the second term in the rhs. We further simplify the problem by assuming that $\eta_I$, $\rho_V$ and the four Stokes components of $\vec{\varepsilon}$ are constant along the slab. With these additional conditions
\begin{equation}
H_I = (s - s'') \eta_I  \, , \quad R_V = (s - s'') \rho_V \, , 
\label{ConstParam}
\end{equation}
and so the evolution operator becomes
\begin{equation}
\hat{O}(s,s'') = \left(\begin{array}{c c c c} 
1 & 0 & 0 & 0 \\
0 & \cos [(s - s'') \rho_V ] & - \sin [(s - s'') \rho_V ] & 0 \\
0 &  \sin [(s - s'') \rho_V ] &  \cos [(s - s'') \rho_V ] & 0 \\
0 & 0 & 0 & 1 \end{array} \right) .
\end{equation}
Thus, the following integrals appear in Eq.~\eqref{EvopRTE}
\begin{subequations}
\begin{align}
T_1 & = \int_{s'}^s \! \mathrm{e}^{- (s - s'') \eta_I} \, \mathrm{d}s'' = \frac{1}{\eta_I} \bigl(1 - \mathrm{exp}^{- (s - s') \eta_I} \bigr) \\
T_2 & = \int_{s'}^s \! \mathrm{e}^{- (s - s'') \eta_I} \cos [(s - s'') \rho_V] \, \mathrm{d}s'' \notag \\
& = \frac{1}{\eta_I^2 + \rho_V^2} \Bigl[ \eta_I - \mathrm{e}^{- (s - s') \eta_I} \Bigl(\eta_I \cos [(s - s') \rho_V] - \rho_V \sin [(s - s') \rho_V] \Bigr)\Bigr] \\
T_3 & = \int_{s'}^s \! \mathrm{e}^{- (s - s'') \eta_I} \sin [(s - s'') \rho_V]  \, \mathrm{d}s'' \notag \\
& = \frac{1}{\eta_I^2 + \rho_V^2} \Bigl[ \rho_V - \mathrm{e}^{- (s - s') \eta_I} \Bigl(\eta_I \sin [(s - s') \rho_V] + \rho_V \cos [(s - s') \rho_V] \Bigr)\Bigr] \, .
\end{align}
\end{subequations}
The Stokes parameters for the radiation emerging from the slab are then given by 
\begin{subequations}
\begin{align}
I(s) & = T_1 \varepsilon_I \, , \quad V(s) = T_1 \varepsilon_Q \\
Q(s) & = T_2 \varepsilon_Q - T_3 \varepsilon_U \\
U(s) & = T_3 \varepsilon_Q + T_2 \varepsilon_U \, ,
\label{StokesEvopSlab}
\end{align}
\end{subequations}
and its linear polarization fraction by
\begin{align}
P_L & = \sqrt{\frac{T_2^2 + T_3^2}{T_1^2}} \sqrt{\frac{\varepsilon_Q^2 + \varepsilon_U^2}{\varepsilon_I^2}} \notag \\
& = \sqrt{\frac{\eta_I^2}{\eta_I^2 + \rho_V^2}} \sqrt{\frac{1 -  2 \mathrm{e}^{- (s - s') \eta_I}  \cos [(s - s') \rho_V] + \mathrm{e}^{- 2 (s - s') \eta_I}}
{1 -  2 \mathrm{e}^{- (s - s') \eta_I} + \mathrm{e}^{- 2 (s - s') \eta_I}}} \sqrt{\frac{\varepsilon_Q^2 + \varepsilon_U^2}{\varepsilon_I^2}}  \, .
\label{PolFracT}
\end{align}
From this last expression the following conclusions can be immediately extracted. Under these assumptions, if the emissivity in the slab is unpolarized, the emerging radiation will also be unpolarized 
independently of the width of the slab and on the value of $\eta_I$ and $\rho_V$. However, provided that the emissivity is polarized, $\rho_V$ can affect the total linear polarization fraction. As could be expected, if $\rho_V$ is zero, the linear polarization fraction will not depend on $\Delta s = s - s'$ or on $\eta_I$, and furthermore, when $\rho_V$ is present, the resulting linear polarization fraction will not depend on its sign. 
\begin{figure*}[htp]
\centering
 \includegraphics[width=16.2cm]{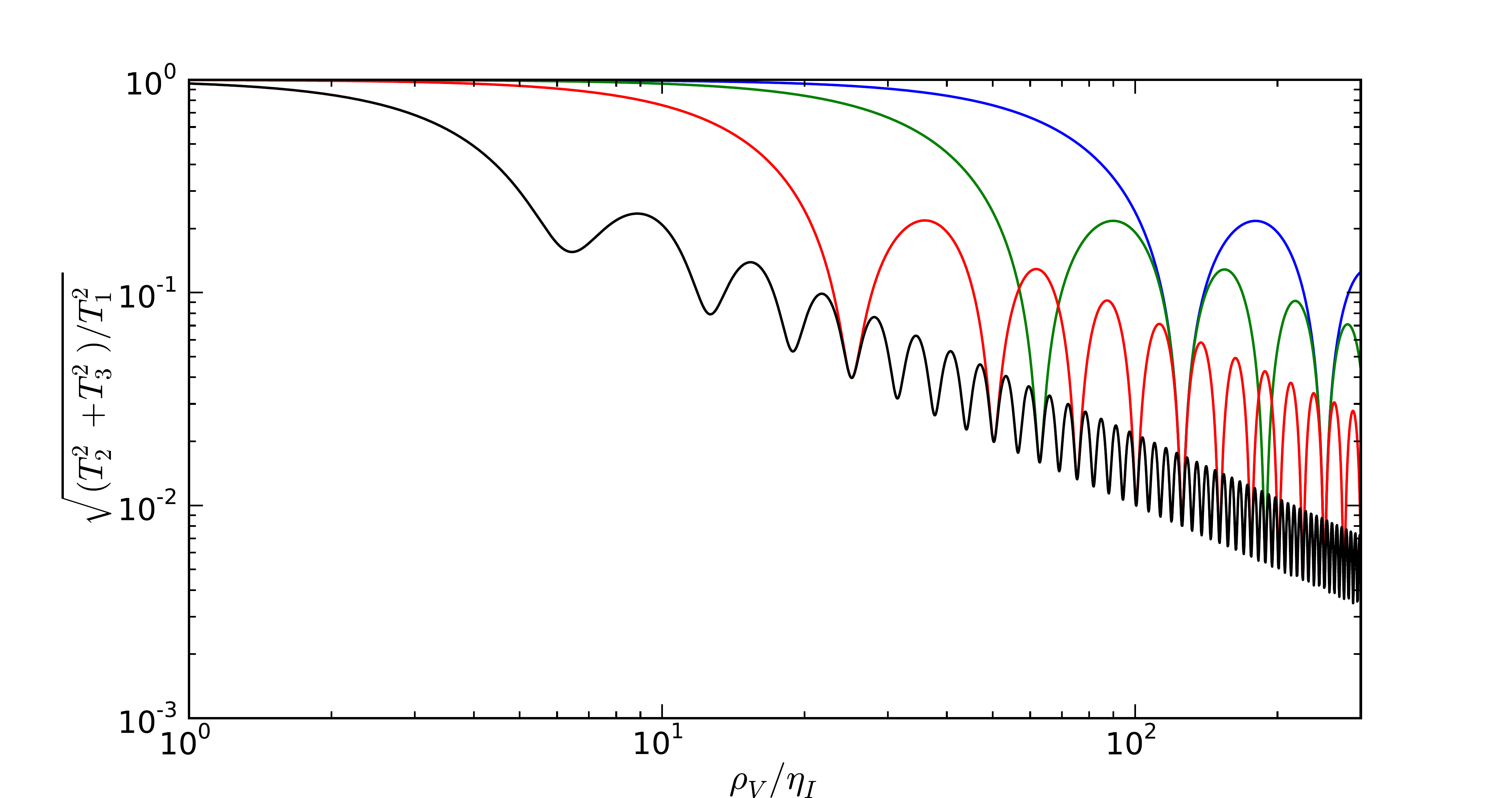}
 \caption{$(T_2^2 + T_3^2)/T_1^2$ factor (see Eq.~\eqref{PolFracT}, as a function of $\rho_V/\eta_I$. The various colored curves correspond to different optical thicknesses of the emitting slab: $\Delta\tau = 1$ (black), $\Delta\tau = 0.25$ (red), $\Delta\tau = 0.1$ (green), and $\Delta\tau = 0.05$ (blue).} 
\label{FigPolFrac}
\end{figure*} 
These magneto-optical effects tend to reduce the linear polarization fraction of the emitted radiation, as shown in Fig.~\ref{FigPolFrac}, where the $(T_2^2 + T_3^2)/T_1^2$ factor is plotted as a function of $\rho_V/\eta_I$. 
 Note that this factor, which is unity in the absence of $\rho_V$, indicates change in the emerging radiation's total linear polarization fraction. We recall that it is assumed that the $\eta_I$, $\rho_V$, and $\varepsilon_i$ are the only nonzero coefficients, and are constant over the whole slab. The various curves in the figure represent different $\Delta s$ for this slab and, given that $\eta_I$ is taken equal to one, they are a measure of different optical thicknesses ranging from $0.05$ to $1$. 
 The linear polarization fraction of the radiation emerging from the slab tends to decrease with $\rho_V$, although it is interesting to note that for certain 
values of $\rho_V/\eta_I$ the rotation of the plane of polarization is particularly efficient in reducing the linear polarization fraction of the emitted radiation. 
Thus, in the plot one finds an oscillatory behavior as a function of $\rho_V/\eta_I$ superimposed on the general decreasing trend. All this is in keeping with the discussion in Sect.~\ref{Sect:MO}, where it is explained how Faraday rotation, operating on the radiation propagating through a finite length of material, may cause a net reduction of the emergent radiation's linear polarization. 
\section{The far-wing behavior of the ${\mathcal R}_{\mbox{\scriptsize II}}$ redistribution matrix}
\label{Sect:WingHanle} 
By definition, scattering processes that are not perturbed by an elastic collision are coherent in the atomic rest frame. Such processes are described by ${\mathcal R}_{\mbox{\scriptsize II}}$, given in \citet{Bommier97b} as  
\begin{align} 
\bigl[{\mathcal R}(\nu',\vec{\Omega}', & \nu,\vec{\Omega};\vec{B}) \bigr]_{ij} = \sum_{K K' Q} \sum_{\substack{ M_u {M_u}' M_\ell {M_\ell}' \\ p p' p'' p'''}} 
{\mathcal C}_{K' K'' Q {M_u} {M_u}' M_\ell {M_\ell}' p p' p'' p'''} \frac{\Gamma_R}{\Gamma_R + \Gamma_I + \Gamma_E + \mathrm{i} \, 2 \pi \nu_L g_u Q} \notag \\
\times & (-1)^Q {\mathcal T}^{K''}_Q(i,\vec{\Omega}) {\mathcal T}^{K'}_{-Q}(j,\vec{\Omega'}) \delta(\nu - \nu' - \nu_{M_\ell, {M_\ell}'})
\frac{1}{2} \biggl[\Phi\bigl(\nu_{{M_u}',M_\ell} - \nu \bigr) + \Phi\bigl(\nu_{{M_u},M_\ell} - \nu \bigr)^\ast \biggr] \, ,
\label{R2}
\end{align}
where $M_u$ and ${M_u}'$ are the quantum numbers of the magnetic sublevels of the upper level; $M_\ell$ and ${M_\ell}'$ are those of the lower level; and $p$, $p'$, $p''$, and $p'''$ are quantum numbers which take integer values between $-1$ and $1$. ${\mathcal C}_{K' K'' Q {M_u} {M_u}' M_\ell {M_\ell}' p p' p'' p'''}$ is a real number which contains the coupling between such angular momentum quantum numbers, and its expression is given in \citet{Bommier97b}. $\nu_L$ is the Larmor frequency and $g_u$ is the Land\'e factor of the upper level. $\Gamma_R$ is the natural broadening of the line, $\Gamma_I$ is the line broadening due to inelastic collisions, and $\Gamma_E$ is the line broadening due to elastic collisions. The total line broadening is given by $\Gamma = \Gamma_R + \Gamma_I + \Gamma_E$. 

In an atmospheric region where the number density of perturbers is high, and thus $\Gamma_E$ is large, it can be seen from the expression of ${\mathcal R}_{\mbox{\scriptsize II}}$ that its branching ratio decreases. Nevertheless, as we have noted in Sect.~\ref{Sect:Dep} and as can be seen from the expression of ${\mathcal R}_{\mbox{\scriptsize III}}$ given in \citet{Bommier97b}, the probability of a photon being emitted at a frequency outside the Doppler core by a line scattering process is much higher if such process is coherent than if it is perturbed by an elastic collision. Thus, the dominant contribution to the line emissivity at wing frequencies comes from ${\mathcal R}_{\mbox{\scriptsize II}}$. We now show that, in the atomic rest frame, the Hanle effect does not modify the polarized radiation emitted at wing frequencies through coherent scattering.

Let us consider the factor in ${\mathcal R}_{\mbox{\scriptsize II}}$ which contains all its magnetic dependence, for each term in the sum over quantum numbers
\begin{equation}
f(\nu',\nu) = \frac{\Gamma_R}{\Gamma_R + \Gamma_I + \Gamma_E + \mathrm{i} \, 2 \pi \nu_L g_u Q}  \delta(\nu - \nu' - \nu_{M_\ell, {M_\ell}'}) \frac{1}{2}
 \biggl[\Phi(\nu_{{M_u}',M_\ell} - \nu) + \Phi(\nu_{{M_u},M_\ell} - \nu)^\ast \biggr] \, .
\label{FR2}
\end{equation}
The profiles $\Phi$ are given, in the atomic rest frame, by
\begin{equation} 
\Phi(\nu_{M_u, M_\ell} - \nu) = \frac{1}{\pi} \frac{1}{\bigl(\Gamma/4 \pi \bigr) - \mathrm{i} \, \bigl(\nu_{M_u, M_\ell} - \nu \bigr)} \, .
\label{ProfLor}
\end{equation}
Accounting for Zeeman splitting
\begin{align*}
\nu_{M_u, M_\ell} & = \nu_0 + \nu_L \, \bigl(g_u M_u - g_\ell M_\ell \bigr) \, ;  \\
\nu_{M_\ell, {M_\ell}'} & = \nu_0 + \nu_L \, g_\ell \, (M_\ell - {M_\ell}') \, ,
\end{align*}
one reaches the relation
\begin{align}
 \frac{1}{2} & \biggl[\Phi(\nu_{{M_u}',M_\ell} - \nu)  + \Phi(\nu_{{M_u},M_\ell} - \nu)^\ast \biggr] = \notag \\
 & = \biggl(1 + \mathrm{i} \frac{2 \pi \nu_L g_u \bigl(M_u - {M_u}' \bigr)}{\Gamma} \biggr) \frac{1}{\pi} \frac{1}{\Bigl[\bigl(\Gamma/4 \pi \bigr) - \mathrm{i} \bigl(\nu_{{M_u}',M_\ell} - \nu \bigr) \Bigr]
  \Bigl[\bigl(\Gamma/4 \pi \bigr) + \mathrm{i} \bigl(\nu_{M_u, M_\ell} - \nu \bigr) \Bigr]}
\label{ProfSum}
\end{align}
Then, considering the far wings, in which $\bigl| \nu_0 - \nu \bigr| \gg \Gamma$ and  $\bigl| \nu_0 - \nu \bigr| \gg \nu_L$, the previous expression becomes
\begin{align}
 \frac{1}{2} & \biggl[\Phi(\nu_{{M_u}',M_\ell} - \nu) + \Phi(\nu_{{M_u},M_\ell} - \nu)^\ast \biggr] \approx \notag \\
 & \approx \biggl(1 + \mathrm{i} \frac{2 \pi \nu_L g_u Q}{\Gamma} \biggr) \frac{1}{\pi} \frac{\Gamma/4 \pi}{\Bigl[\bigl(\Gamma/4 \pi \bigr)^2 + \bigl(\nu_0 - \nu \bigr)^2 \Bigr]} \, ,
\label{ProfSum2}
\end{align}
where $Q = M_u - {M_u}'$. Thus, the factor in Eq.~\eqref{FR2} becomes
\begin{equation}
f(\nu',\nu)  = \frac{\Gamma_R}{\Gamma} \frac{1}{\pi} \frac{\bigl(\Gamma/4 \pi \bigr)}{\bigl(\Gamma/4 \pi \bigr)^2 + \bigl(\nu_0 - \nu \bigr)^2} \, \delta\bigl(\nu - \nu' - \nu_L \, g_\ell \, (M_\ell - {M_\ell}')\bigr) \, .
\label{FR2b}
\end{equation}
The magnetic dependence contained in the Hanle depolarization factor is canceled in the far wings by the magnetic dependence in the Lorentzian profiles. Thus, the Hanle effect does not operate at such frequencies. Nevertheless, the Dirac delta function relating the frequencies of the incoming and scattered radiations is still sensitive to the splitting of the magnetic sublevels of the lower level. We note, however, that the lower level of the Ca{\sc i} $4227$~\AA\ line has $J_\ell = 0$, so ${\mathcal R}_{\mbox{\scriptsize II}}$ has no magnetic dependence in the far wings at all.

\bibliographystyle{apj}

\end{document}